# Synthesis and self-assembly of aminyl and alkynyl substituted sophorolipids

Abdoul Aziz Ba,[a] Jonas Everaert,[b] Alexandre Poirier,[a] Patrick Le Griel,[a] Wim Soetaert,[c] Sophie L. K. W. Roelants,[c] Daniel Hermida-Merino,[d] Christian V. Stevens,[b] Niki Baccile[a,*]

[a] Sorbonne Université, Centre National de la Recherche Scientifique, Laboratoire de Chimie de la Matière Condensée de Paris, LCMCP, F-75005 Paris, France

[b] SynBioC, Department of Green Chemistry and Technology, Ghent University, Ghent, Belgium

[c] InBio, Department of Biotechnology, Ghent University, Ghent, Belgium

[d] Netherlands Organisation for Scientific Research (NWO), DUBBLE@ESRF BP CS40220, 38043 Grenoble, France

**\* Corresponding author:**
Dr. Niki Baccile
E-mail address: niki.baccile@sorbonne-universite.fr
Phone: +33 1 44 27 56 77

**Abstract**

Sophorolipids are one of the most important microbial biosurfactants, because of their large-scale production and applications developed so far in the fields of detergency, microbiology, cosmetics or environmental science. However, the structural variety of native sophorolipids is limited/restricted, a limiting fact for the development of new properties and their potential applications. In their open acidic form, C18:1 sophorolipids (SL) are classically composed of a sophorose headgroup and a carboxylic acid (COOH) end-group. The carboxyl group gives them unique pH-responsive properties, but they are a poorly-reactive group and their charge can only be negative. To develop a new generation of pH-responsive, positively-charged, SL and to improve their reactivity for further functionalization, we develop here SLs with an amine (-NH$_2$) or terminal alkyne (-C≡CH) end-group analogues. The amine group generates positively-charged SL and is more reactive than carboxylic acids, e.g. towards aldehydes; the alkyne group provides access to copper-based click chemistry. In this work, we synthesize (C18:1) and (C18:0) –NH$_2$ and (C18:1) -C≡CH sophorolipid derivatives and we study their self-assembly properties in response to pH and/or temperature changes by means of static and dynamic light scattering, small angle (X-ray, neutron) scattering and cryogenic electron microscopy. Monounsaturated aminyl SL-C18:1-NH$_2$ sophorolipids form a micellar phase in their neutral form at high pH and a mixed micellar-bilayer phase in their positively-charged form at low pH. Saturated aminyl SL-C18:0-NH$_2$ sophorolipids form a micellar phase in their charged form at low pH and a twisted ribbon phase in their neutral form at high pH and monounsaturated alkynyl SL-C18:1-C≡CH sophorolipids form a main micellar phase at T> 51.8°C and a twisted ribbon phase at T< 51.8 °C.

**Introduction**

Microbial biosurfactants are one of the most interesting alternatives to replace petrochemical surfactants.[1,2] They are produced from renewable resources of both first and second generation[3,4] and they are broadly considered as minimally eco- and cytotoxic compounds.[5] Their unique molecular structure, often composed of two asymmetric hydrophilic end-groups, promotes a rich phase behavior: micellar,[6–8] fibrillar,[9] lamellar,[10,11] vesicular,[10,12] or sponge[11] structures have been reported in the literature over the past fifteen years. Mastering the biosurfactants' self-assembly is crucial to make both soft[13–16] and inorganic[17] materials, with unexpected properties.[14,15] Strong lamellar and fibrillar hydrogels with stimuli responsivity can be prepared respectively from gluco- and sophorolipids;[14,18–20] gene transfection capsules can be prepared by integrating sophorolipid derivatives[21] or MELs[16] to lipid vesicles; self-standing lamellar foams with Young moduli up to 30 kPa,[15] bulk or surface antimicrobial and antiadhesive coatings,[22–25] capped nanoparticles[26–29] with narrow size distribution and high colloidal stability[26] are just some of the most prominent examples recently developed.

Microbial biosurfactants have two main limitations besides their production costs.[30] Low purity (residual fatty and organic acids) and uniformity (congener variety) are responsible for poor batch homogeneity, which can affect the phase behavior,[31] consequently limiting the application of commercial formulations, of which the composition may vary from batch to batch. In the meantime, the structural variety of the most abundant molecular forms is limited, thus containing the diversity in properties and limiting the application potential. The latter is a well-known issue, addressed since at least three decades,[32–36] but strongly pushed forward by a strong contribution from organic chemistry,[35–39] by recent developments in the fields of genetic engineering[40–43] and by the combination of both approaches.[10,44]

Genetically modified *Starmerella bombicola* yeasts, of which the wild type produces the well-known C18:1 sophorolipid, produce variants such as single-glucose lipids (glucolipids),[40] C16:0,[41] symmetrical and asymmetrical bola[45] or C18:1 acetylated sophorolipids.[46] Such structural variety shifts the self-assembly properties from classical micellization obtained with standard acidic C18:1 sophorolipids,[47] to vesiculation with[48] or without[10] encapsulation potential and hydrogel-formation[20] potential. Nonetheless, genetic engineering takes long time and efforts to develop and yields may be small, while chemical derivatization can virtually provide an infinite variety of molecules.[37] In this regard, sophorolipid esters were synthesized in view of improving the emulsifying properties of acidic sophorolipids,[39,49,50] quaternary ammonium sophorolipid derivatives were developed as transfection carriers[21,51] or to improve the antimicrobial effects of sophorolipids.[51] Finally, the combination of the genetic engineering and chemical derivatization is also interesting, as shown by the development of saturated compounds.[9,10,52] Well-known in oil science, saturation introduces a rigidity in the lipid chain, it improves the molecular order and increases the melting temperature. These features were used within the framework of sophoro and glucolipids to develop hydrogels and foams.[10,14,19]

The carboxylic COOH end-group is certainly one of the most interesting features of many glycolipid biosurfactants, as it modulates the lipid charge, also being an interesting site to perform chemical modifications. However, carboxylic acids only provide negative charges and their reactivity is energetically unfavored, whereas efficient derivatization of organic acids requires the prior formation of the more reactive esters. For this reason, we develop two new sophorolipids derivatives with amine (-NH$_2$) and alkyne (-C≡CH) functions. The amine is interesting for its stimuli (pH) responsiveness, similarly to carboxylic acids, but providing an opposite, positive, charge. The prompt reactivity of amines towards aldehydes in aqueous environment is also an interesting aspect for further structural modifications. In this regard, we also develop alkynyl sophorolipid derivatives, interesting for aqueous Cu(I)-catalyzed azide-alkyne cycloaddition (CuAAC) click reactions.[53] Knowing the self-assembly properties of aminyl and alkynyl derivatives of sophorolipids in water could lead to the development of fully-sustainable glyconanomaterials.[54–56] Amynil and alkynyl end-groups could be used in biosurfactant-based[10,12] gene transfection capsules,[57] needing a globally positive surface charge but also possible gene/antigene or fluorescent coupling, the latter being a straightforward task with alkynyl functions. Mixing sophorolipids with carboxylic and amine functions could lead to the development of new pH-responsive catanionic biosurfactant systems, where new phases could be developed through the precise control of positive/negative charge ratio, strongly influencing the local curvature[58,59] and with applications in cosmetics, pharmacy or pollution control.[60] The latter field is particularly interesting. If negatively-charged biosurfactants have long been used for heavy metal ion removal,[61] positively charged biosurfactants could be employed for removal of phosphates and nitrates, responsible for coastal water eutrophication,[62,63] while more complex aminyl/alkynyl formulations could combine anion removal with selective metal depollution, whenever the alkyne group would be modified with crown ethers or cyclodextrins.

In this work are synthesized both C18:1 and C18:0 aminyl and C18:1 alkynyl sophorolipids (SL) and were characterized their self-assembly in water under dilute conditions combining small angle X-ray scattering (SAXS), small-angle neutron scattering (SANS), static and dynamic light scattering and transmission electron microscopy under cryogenic conditions (cryo-TEM). The pH-responsiveness of monounsaturated SL-C18:1-NH$_2$ and saturated SL-C18:0-NH$_2$ sophorolipids were evaluated and compared to such behavior of their acidic C18:1 (SL-C18:1-COOH) and C18:0 (SL-C18:0-COOH) sophorolipid counterparts, largely studied in the past. The same study is performed on C18:1 alkynyl sophorolipids (SL-C18:1-C≡CH), of which the self-assembly was studied against temperature.

**Experimental Section**

*Chemicals*. Lactonic sophorolipids are constituted of 93.4% C18:1 diacetylated subterminal hydroxylated sophorolipid lactone and are produced in previous studies.[57,64] Ethylenediamine and propargylamine are purchased at Aldrich and used as such. The carbon-supported palladium (Pd/C) and Novozym 435 are purchased at Sigma-Aldrich. Figure 1 shows the structure of the monounsaturated and saturated aminyl and alkynyl sophorolipids used in this work. Their synthesis procedure, $^1$H and $^{13}$C NMR analyses are detailed in the Supporting Information, Page S2-S7 and Figure S 1 to Figure S 4.

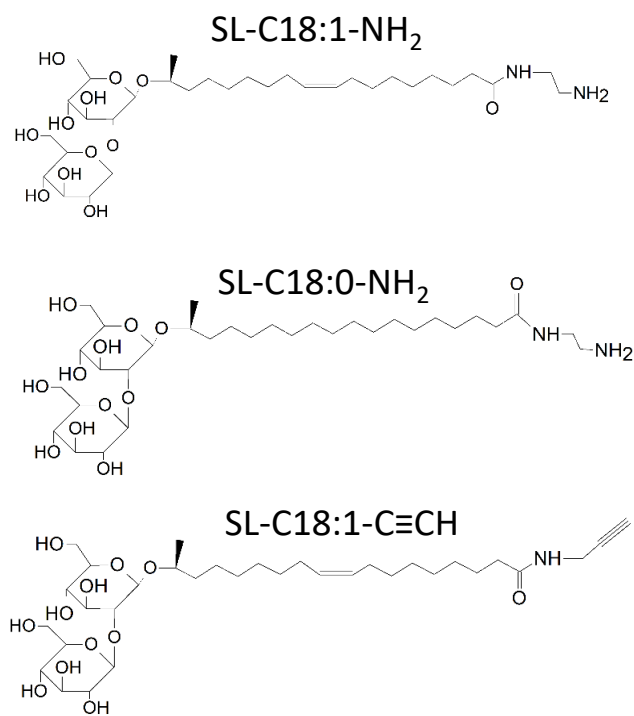

**Figure 1** – Chemical structure of monounsaturated aminyl sophorolipid, SL-C18:1-NH$_2$, saturated aminyl sophorolipid, SL-C18:0-NH$_2$ and monounsaturated alkynyl sophorolipid, SL-C18:1-C≡CH.

*Static and dynamic light scattering (LS)*. Static and dynamic LS experiments are performed using a Malvern Zetasizer Nano ZS90 (Malvern Instruments Ltd, Worcestershire, UK) equipped with a 4 mW He–Ne laser at a wavelength of 633 nm. Unless otherwise stated, measurements were made at 25 °C. All experiments are done at a fixed angle of 90°. For temperature-dependent experiments, the temperature is controlled through the Zetasizer nano software, allowing 240 s of equilibration time between two temperatures. For the experiments involving the evolution of the scattered light with concentration (critical micelle concentration experiments), we consider the mean count rate at a constant value of the attenuator (11, equivalent to no attenuation of the signal).

*Turbidimetric titration using Light Scattering (LS) and ζ-potential.* Turbidimetric and ζ-potential experiments are performed using the automatic titration unit MPT-2 of a Malvern Zetasizer Nano ZS90 (Malvern Instruments Ltd, Worcestershire, UK) instrument, equipped with a 4 mW He-Ne laser at a wavelength of $\lambda = 633$ nm, measuring angle, $\theta = 90°$, temperature, $T = 25°C$, and the signal is never attenuated throughout the entire experiment. The sample solution ($V = 8$ mL) is contained in an external beaker and pumped with a peristaltic pump through the ζ-potential cuvette cell located in the instrument for analysis. Equilibrium pH of each sample solution is about 8.6 and it is increased to about pH 11 by adding 1 µL of a 5 M NaOH solution. pH is adjusted in the beaker by adding aliquots of $V = 5$ µL of a HCl solution at $C = 0.5$ M and controlled by the MPT-2 Zetasizer software. pH is measured *in situ* by coupling a calibrated KCl electrode to the MPT-2 unit. The beaker undergoes gentle stirring to avoid the formation of air bubbles in the flow-through tubing system and, consequently, in the ζ-potential cuvette. Avoiding air bubbles in the cuvette is crucial and accurately inspected throughout the experiment. Light scattering and ζ-potential are simultaneously recorded between each pH variation while the sample solution is continuously pumped through the cuvette.

*Differential Scanning Calorimetry (DSC)*: DSC is performed using a DSC Q20 apparatus from TA Instruments equipped with the Advantage for Q Series Version acquisition software (v5.4.0). Acquisition is performed on a dry powder sample (~ 10 mg) sealed in a classical aluminium cup and using a heating ramp at a rate of $10°C.min^{-1}$.

*Cryogenic Transmission Electron Microscopy (Cryo-TEM).* These experiments were carried out on an FEI Tecnai 120 twin microscope operating at 120 kV equipped with a Gatan Orius CCD numeric camera. The sample holder was a Gatan Cryoholder (Gatan 626DH, Gatan). Digital Micrograph software was used for image acquisition. Cryofixation was done on a homemade cryofixation device. The solutions were deposited on a glow-discharged holey carbon coated TEM copper grid (Quantifoil R2/2, Germany). Excess solution was removed and the grid was immediately plunged into liquid ethane at −180 °C before transferring them into liquid nitrogen. All grids were kept at liquid nitrogen temperature throughout all experimentation.

*Scanning Electron Microscopy with Field Emission Gun* (SEM-FEG): SEM-FEG experiments have been recorded on a Hitachi SU-70. The images were taken in secondary electron mode with an accelerating voltage at 1 kV, 5 kV or 10 kV. Prior to analysis, the materials were coated with a thin layer of gold by sputter deposition.

*Scattering experiments*: Small angle X-ray and neutron scattering (respectively, SAXS and SANS) experiments and analysis of SAXS/SANS data are described in detail in the Supporting Information, Page S8-S11 and in Figure S 5.

**Results and discussion**

*Aminyl sophorolipid.* SL-C18:1-NH$_2$ is prepared by reacting ethylene diamine with deacetylated acidic C18:1 sophorolipids. The synthesis was tested employing two approaches, the use of Novozym 435 at mild temperature (50°C) in THF and the use of high temperature (120°C) in methanol, where the enzymatic approach is known to efficiently catalyze amidation reactions under milder conditions.[33,35] Unfortunately, this approach provided the formation of dimeric sophorolipids species, which were not desired in this work. For this reason this approach was abandoned. On the contrary, the methanolic synthesis revealed to be more efficient, with a yield of 84% of the amide and the absence of sophorolipid dimers. $^1$H NMR spectroscopy (Figure S 3) shows a set of peaks typical for sophorolipid compounds,[33,57,65] in particular the triplet above 2.1 ppm, reflecting the CH$_2$ in α position of C=O (peak 2 in Figure S 3), the CH=CH around 5.34-5.38 (peak 9,10 in Figure S 3) and the consequent peak reflecting positions 8,11 (Figure S 3) between 2.02-2.08 ppm, as well as anomeric protons of sophorose between 4.45 and 4.64 ppm (peaks 1´, 1´´ in Figure S 3). Proof of functionalization and amide formation is given by peaks at 2.89 ppm attributed to the CH$_2$NH$_2$ (peak 20 in Figure S 3), by the upfield shift of the CH$_2$CONH (position α with respect to C=O, peak 2 in Figure S 3) from the classical value of 2.30 ppm found in acidic (COOH) sophorolipids[65,66] to 2.19 ppm in amide derivatives of sophorolipids.[33] Attribution of the CONHCH$_2$ group is less precise, due to the overlap with the (6x)CHOC groups of sophorose in the range 3.21-3.41 ppm (peaks 21 in Figure S 3). $^{13}$C peaks at 40.4 (CONHCH$_2$) and 41.4 (CH$_2$NH$_2$) ppm confirm the $^1$H spectra, while relative integration of the $^1$H peaks, in particular close-to 1:1 ratio between CH$_2$NH$_2$ and CH$_2$CONH groups confirm single grafting and it excludes the formation of dimers. The resulting SL-C18:1-NH$_2$ compound after column chromatography is obtained in the form of a yellowish sticky melt at room temperature.

SL-C18:1-NH$_2$ is water soluble in the concentration range explored in this work (below 1 wt%) and at its equilibrium pH (~8.6). The slight alkaline pH and the positive value of the electrophoretic mobility at this pH (Figure S 6b) confirm the successful amine modification, considering that the equilibrium pH of deacetylated acidic sophorolipids, SL-C18:1-COOH, is generally between 4 and 5 within a similar concentration range. In analogy with SL-C18:1-COOH, the solubility of SL-C18:1-NH$_2$ most likely depend on the formation of a micellar phase. Concentration-dependent static light scattering experiments (Figure S 7) show a concentration above which light is scattered, this being the typical behavior for surfactant solutions above their critical micelle concentration (cmc), estimated at 0.5 ± 0.2 mg/mL for SL-C18:1-NH$_2$ (Figure S 7a). This cmc is in the same order found for SL-C18:1-COOH sophorolipids,[67] thus showing that the presence of the amine does not change drastically this specific property.

The micellar phase and structure are studied in more detail by SAXS and SANS experiments. Figure 2b shows three SAXS profiles recorded on an aqueous SL-C18:1-NH$_2$ solution at $C$= 0.5 wt% and three pH values, from basic to acidic. For all pH, the SAXS data display the typical profiles of a micellar form factor, as found for acidic SL-C18:1-COOH sophorolipids.[8] All SAXS profiles show a plateau between 0.2 and 0.5 nm$^{-1}$, indicating the finite size of the micelles in the $q$-range explored in this work. Above 1 nm$^{-1}$, the profiles show the first oscillation of the form factor, with a minimum at about 2 nm$^{-1}$. We analyze the SAXS data with a model-independent and a model-dependent approach. The finite size of the micelles allows a safe use of the Guinier approximation (please refer to the materials and method section for more details) in the range $q \cdot R_g < 1$, where $R_g$ is the radius of gyration. The linear fit of the $Log(I)$ vs. $q^2$ representation in the portion below 0.5 nm$^{-1}$ (Figure 2c) gives $R_g$ for the three pH conditions. The fit parameters, as well as the $R_g$ values, are summarized in Table 1, whereas the relationship between $b$ (slope of the linearized Guinier expression), $R_g$ and $R_{Guinier}$ are given in the Supporting Information (Page S9). The radius of the micellar aggregates obtained from the Guinier approximation, $R_{Guinier}$, is contained between ~2 and ~2.5 nm, a range of values in good agreement with the values reported for acidic sophorolipids, varying from 1.5 nm to 2.5 nm according to pH.[68] This range is also slightly smaller than a fully elongated sophorolipid molecules, of which the size is rather expected in the order of 3 nm, thus suggesting a possible bent configuration of the lipid within the micelles.

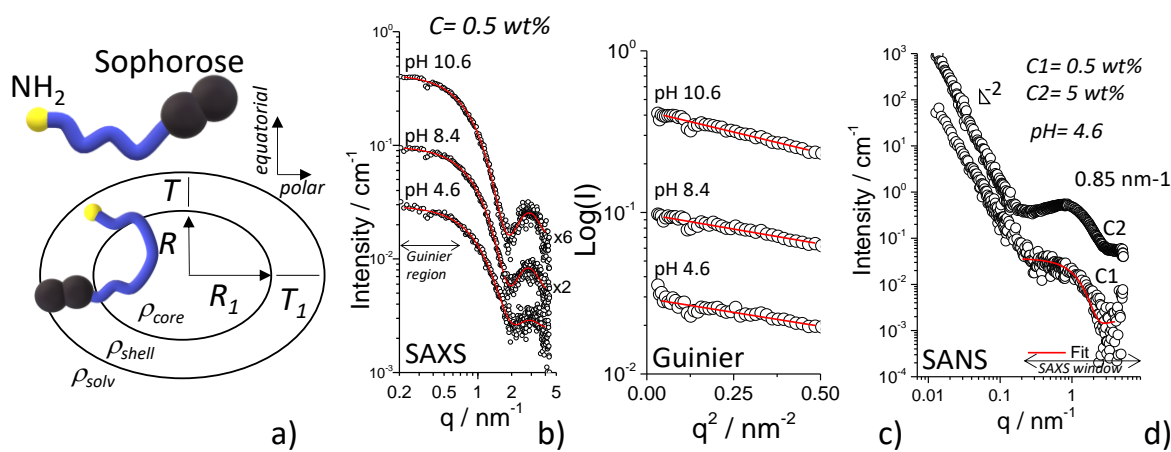

**Figure 2 – Small angle experiments recorded on SL-C18:1-NH₂ solutions. a)** Core-shell *"coffee-bean"*-like form factor model used to fit SAXS and SANS data. **b)** SAXS profiles at $C$ = 0.5 wt% and pH between 4.6 and 10.6. **c)** Guinier plots of SAXS profiles in b) within the $qR_g < 1$ approximation. **d)** SANS profiles recorded at pH 4.6 and concentrations of 0.5 wt% and 5 wt%.

If the Guinier approximation provides a realistic estimation of the micellar radius, it cannot explain more peculiar features of the SAXS profiles in Figure 2b, and in particular the remarkable increase in the amplitude of the form factor oscillation above 1 nm$^{-1}$ from acidic to basic pH. The amplitude is very sensitive to the contrast at the solvent/shell/core palisade and its evolution is a strong hint of a variation of the electron densities with pH. Similar features were found in the SAXS patterns of SL-C18:1-COOH, but where the amplitude increases in the opposite sense, from basic to acidic pH.[68] We have previously shown that micellar aggregates of SL-C18:1-COOH sophorolipids can be described by an atypical *"coffee-bean"* model, corresponding to a core-shell (prolate) ellipsoid of revolution, of which the shell thickness is not homogeneous in the equatorial and polar regions of the micelle,[8] as otherwise assumed in most surfactant micelles.[69] The *"coffee-bean"* model is sketched in Figure 2a and presented in detail in the Supporting Information (Page S9-S11, Figure S 5).

The *"coffee-bean"* model is obviously an approximation and it is physically unrealistic *per se*; however, it is the best model that we have found to precisely match the SAXS oscillation of the form factor across the entire pH range in sophorolipid systems,[8] whereas a simpler model (core surrounded by a homogeneous shell) fails. The *"coffee-bean"* model should be understood in terms of an inhomogeneous electron density distribution at the core-shell interface for this class of bolaform amphiphiles, of which the carboxylic acid end-group can be more or less hydrophilic according to its protonation state and, consequently, occupy different regions within the micelle.[8]

**Table 1 – Numerical results obtained from the analysis of the Guinier plots in Figure 2c.** $b$ $(= -0.434 \frac{R_g^2}{3})$ **corresponds to the slope of the linearized Guinier expression and it is defined in the Supporting Information (Page S9).**

| pH | $b$ | R-squared | $R_g$ | $R_{Guinier}$ |
|---|---|---|---|---|
| 4.6 | -0.345 ± 0.027 | 0.952 | 1.54 ± 0.06 | 1.99 ± 0.08 |
| 8.6 | -0.368 ± 0.021 | 0.900 | 1.59 ± 0.05 | 2.06 ± 0.06 |
| 10.6 | -0.515 ± 0.020 | 0.823 | 1.89 ± 0.04 | 2.44 ± 0.05 |

The use of the *"coffee-bean"* model to fit the SAXS data in Figure 2b is justified by the close structural similarity between SL-C18:1-NH$_2$ and SL-C18:1-COOH sophorolipids but, above all, by their very similar SAXS patterns and, very interestingly, by the inverse evolution of the oscillation amplitude against pH, increasing with increasing pH for SL-C18:1-NH$_2$ (Figure 2b) and decreasing with increasing pH for SL-C18:1-COOH.[68] Table 2 reports the typical equatorial shell thickness, $T$, and core radius, $R$, as well as the shell ($\frac{T_1}{T}$) and core ($\frac{R_1}{R}$) aspect ratios, obtained from fitting the SAXS patterns in Figure 2b using the "*coffee-bean*" micellar model (Figure 2a). Table 2 also compares the structural micellar parameters obtained for SL-C18:1-NH$_2$ (this work) with the ones reported for SL-C18:1-COOH.[68] As expected, the SAXS profiles of SL-C18:1-NH$_2$ can be satisfactorily fitted with a non-homogeneous shell ($\frac{T_1}{T} \neq 1$), where the core aspect ratio $\frac{R_1}{R} \neq 1$ indicates that micelles are ellipsoids and not spheres. Each parameter is addressed independently below:

- *Equatorial core radius*. $R$ lies around 0.7 (± 0.1) nm and it is practically unchanged with respect to pH, in agreement with the findings for acidic sophorolipids SL-C18:1-COOH. This shows that the micellar core in sophorolipid micelles, despite their pH sensitive end-chemical group (-NH$_2$ or -COOH), is quite narrow and poorly affected by the physicochemical environment.

- *Core aspect ratio*. The polar dimension of the core, $R_1$, is definitely much more affected by the end-group and pH. Micelles are almost systematically described as prolate ellipsoids of revolution, because $\frac{R_1}{R}$ is generally greater than 1. Interestingly, $\frac{R_1}{R}$ tends to unity upon ionization of the end-group: at pH 4.6 for SL-C18:1-NH$_2$ and at pH 10.5 for SL-C18:1-COOH, respectively characterized by the protonated $NH_3^+$ and $COO^-$ forms. pH-dependent electrophoretic mobility experiments (Figure S 6b) of the former confirm an overall positive charge below pH 10. Such a structural similarity ($\frac{R_1}{R} \rightarrow 1$) between the ammonium and carboxylate sophorolipids can be explained by the presence of repulsive electrostatic forces, being the driving force for a higher curvature and more spherical micelles.

- *Equatorial shell thickness*. $T$ is in the order of 1 nm for SL-C18:1-NH$_2$, a value which is compatible with the size of a disaccharide.[70] $T$ seems quite independent of pH, differently than what it is observed for SL-C18:1-COOH, where values of $T$ in the order of few Ångstrom could be used to fit the data upon ionization into carboxylates. It is however unclear, at the moment, whether such discrepancy in terms of $T$ between the amine and carboxylic sophorolipids is physically meaningful or driven by the uncertainties of the fitting strategy using the "*coffee-bean*" micelle model.

- *Thickness aspect ratio*. The thickness $T_1$ in the polar direction of the ellipsoid is very sensitive to pH and this is observed both for amine and carboxylic sophorolipids, where $\frac{T_1}{T} > 1$ is systematically observed for the ionized forms. Similarly, $\frac{T_1}{T} < 1$ is observed for the neutral micellar environment obtained with both compounds.

Table 2 – Numerical values of the structural parameters obtained from the fit of SAXS data associated to SL-C18:1-NH₂ (Figure 2b) and SL-C18:0-NH₂ (Figure 5) samples using the "*coffee-bean*" model presented in Figure 2a and discussed in the Supporting Information (Page S9-S11). The values given for the SL-C18:1-COOH and SL-C18:0-COOH samples are extracted from Ref. [68] and the pH given in the table.

| Compound | pH | $T$ / nm | $R$ / nm | $T_1/T$ | $R_1/R$ | $\rho_{shell}$ / $10^{-4}$ nm$^{-2}$ |
|---|---|---|---|---|---|---|
| | | | C18:1 congener | | | |
| SL-C18:1-NH₂ | 4.6 | 1.08 | 0.67 | 1.8 | 0.8 | 10.5 |
| | 8.5 | 0.84 | 0.85 | 1.3 | 1.6 | 11.2 |
| | 10.6 | 1.01 | 0.74 | 0.1 | 4.2 | 11.4 |
| SL-C18:1-COOH[68] | 4.5 | 1.25 | 0.75 | 0.5 | 4.0 | 10.7 |
| | 8.5 | 0.30 | 0.80 | 3.5 | 1.7 | 11.2 |
| | 10.5 | 0.30 | 0.85 | 3.5 | 1.7 | 11.3 |
| | | | C18:0 congener | | | |
| SL-C18:0-NH₂ | 4.6 | 0.60 | 0.70 | 3.5 | 1.5 | 9.91 |
| | 8.5 | | | Fiber phase | | |
| SL-C18:0-COOH[68] | 4.6 | | | Fiber phase | | |
| | 8.5 | 0.60 | 0.90 | - | 1.4 | 10.7 |

SAXS experiments demonstrate the presence of a SL-C18:1-NH₂ micellar phase across a wide pH range from acidic to basic. The model-independent Guinier approximation indicates a micellar radius ($R_{Guinier}$) between ~2 nm and ~2.5 nm, in good agreement with the total equatorial $R_{CB\ model}\ (= R + T)$ and polar $R_{1\ CB\ model}\ (= R_1 + T_1)$ micellar radii obtained from the "*coffee-bean*" model, all given in Table 3 for direct comparison. However, the model-dependent approach provides a refined picture of the local micellar structure. SL-C18:1-NH₂ micelles have a small hydrophobic core, which is roughly spherical upon ionization into $NH_3^+$ at pH 4.6 and which elongates in one direction up to a factor 4 in its NH₂ form above pH 10. Electrophoretic mobility data (Figure S 6b) corroborate the evolution from charged to neutral colloids when increasing pH. The shell thickness has an opposite behavior; upon ionization (pH 4.6), the thickness in the polar direction stretches up to a factor 2, while it becomes very narrow upon neutralization above pH 10. Interestingly, the same exact trend is observed for the classical acidic sophorolipids, SL-C18:1-COOH, although for inverse pH values, due to the obvious opposite response towards pH of the amino and carboxylic chemical groups.

The similar trend for these two compounds generalizes the understanding of the micellar structure of asymmetric bolaform pH-responsive glycolipids. In the neutral (COOH or NH$_2$) form, sophorolipids self-assemble into micelles with a prominent ellipsoidal core, where the core aspect ratio can reach values as high as 4, and where the hydrophilic groups are mainly located in the equatorial shell region. However, upon ionization towards either $COO^-$ or $NH_3^+$, incoming electrostatic repulsive forces increase the micellar curvature, thus driving a transition from ellipsoids to spheres, classical in ionic surfactants.[71,72] It is however uncommon that the ellipsoid-to-sphere transition only occurs in the hydrophobic core, while the hydrophilic species undergo an uneven redistribution in the shell, being more prominent in the polar than in the equatorial direction.

Complementary SANS experiments were performed in the $NH_3^+$ form of SL-C18:1-NH$_2$ at pH 4.6 and for a broader $q$-range. The generally-admitted advantage of SANS with respect to SAXS in soft colloidal systems is the better contrast between the deuterated solvent and the hydrogenated micelle and which could possibly result in a better description of the micellar structure. The SANS profile of SL-C18:1-NH$_2$ at pH 4.6 is shown in Figure 2d at two concentrations, 0.5 wt% and 5wt%. The $q$-region between 0.2 and 5 nm$^{-1}$, analogous to SAXS, is characterized by a broad signal, where the oscillation of the form factor is now lost. This is not uncommon, because the scattering profile is strongly affected by the difference in the density contrast between the solvent and the micelles with respect to neutrons but also on the level of incoherent signal due to the hydrogen atoms in the system. The combination of both issues generate a less-resolved scattering profile in SANS than in SAXS experiments. Nonetheless, it is possible to fit the SANS profile at 0.5 wt% with exactly the same form factor model and structural values employed in the fit of the corresponding SAXS profile (Table 2). The solvent and core scattering length densities (SLD) used in the fit of SAXS data are replaced with the corresponding values expected for neutron scattering (6.3×10$^{-4}$ nm$^{-2}$ for D$_2$O and 2×10$^{-5}$ nm$^{-2}$ for an aliphatic chain) and for an optimized shell SLD of 4.9×10$^{-4}$ nm$^{-2}$, which is in agreement with the SLD of a carbohydrate undergoing a H/D exchange of its labile protons.[73] The fact that SANS data can be fitted using the form factor model with exactly the same set of structural parameters and adapted SLD strengthens the results obtained from SAXS.

**Table 3** – Comparison between the radii obtained from the model-independent (Guinier) and model-dependent ("*coffee-bean*", CB model) analyses of the SAXS data for SL-C18:1-NH$_2$ solutions at 0.5 w%. $R_{CB\ model} = R + T$, with $R$ and $T$ respectively being the equatorial core radius and shell thickness and $R_{1\ CB\ model} = R_1 + T_1$, with $R_1$ and $T_1$ being the polar core radius and shell thickness. $R, T, R_1, T_1$ are given in Table 2 and described in Figure 2a.

| pH | $R_{Guinier}$ / nm | $R_{CB\ model}$ | $R_{1\ CB\ model}$ |
|---|---|---|---|
| 4.6 | 1.99 ± 0.08 | 1.75 ± 0.18 | 2.48 ± 0.25 |
| 8.5 | 2.06 ± 0.06 | 1.69 ± 0.17 | 2.45 ± 0.25 |
| 10.6 | 2.44 ± 0.05 | 1.75 ± 0.18 | 3.21 ± 0.32 |

SANS experiments show however an interesting feature in the $q$-range out of the SAXS window. Below 0.2 nm$^{-1}$ the signal is characterized by a strong scattering with slope in the log-log plot of -2.6 ± 0.1 at 0.5 wt% and -3.2 ± 0.1 at 5 wt%. The overall SANS signal of the ionized form of SL-C18:1-NH$_2$ is in fact very similar to the ionized form of SL-C18:1-COOH, presented elsewhere.[47,74] In that case, the intense low-$q$ scattering was attributed to the presence of a coexisting nanoplatelet phase.[74] The slopes found here cannot be associated to a specific morphology but they rather suggest a fractal system, classical for non-integer slope values.[75]

The corresponding cryo-TEM experiments at pH 4 presented in Figure S 8 corresponding to the ionized, $NH_3^+$ form, form of aminyl sophorolipids show two distinct families of self-assembled structures.

- *Micellar aggregates* can be observed at high magnification in the vitrified water holes throughout the grid. They are characterized by nanometer-sized spheroidal objects, generally hard to distinguish without ambiguity with our cryo-TEM apparatus. However, micelles can be identified indirectly within the aggregates. Figure S 8c shows a typical image of micellar aggregates, of which the Fourier Transform (FT) panels (FT1 – FT3) on the right-hand side show a broad correlation peak, typically observed for interacting objects. The average distance corresponding to the middle of the ring is 4 ± 1 nm, in very good agreement with a system composed of micelles with $2R_{Guinier}$ = 3.98 ± 0.16 nm and undergoing electrostatic repulsion due to the presence of the ammonium function.

- *Bilayer fragments* constitute the second phase. In cryo-TEM, they often display as cloudy aggregates of several microns (Figure S 8a,b), similarly to what was found for ionized acidic sophorolipids at pH above 10. A closer look within the aggregates (Figure 3a) shows the coexistence of poorly contrasted, flat, and highly contrasted, needle-like, morphologies, the latter being spotted with white arrows numbered 1 to 5. The same image recorded at a +30° tilt angle of the goniometer head of the microscope (Figure 3b) excludes any fiber morphology and demonstrates that needles correspond to the side of bilayers, because they become less contrasted upon tilting (segmented arrows 1-5 in Figure 3b). This is highlighted for arrows 3 and 4, respectively in Figure 3c,d (0°, +30°) and Figure 3e,f (0°, +30°). The thickness of the bilayers varies between 5 and 8 nm, as suggested by the thickness profiles recorded orthogonally to each numbered membrane (Figure 3g). Interestingly, this range of values is consistent with the size domain identified by the broad correlation peak centered at $q= 0.85$ nm$^{-1}$ (7.4 nm) in the SANS profile for the sample recorded at 5 wt% (Figure 2c).

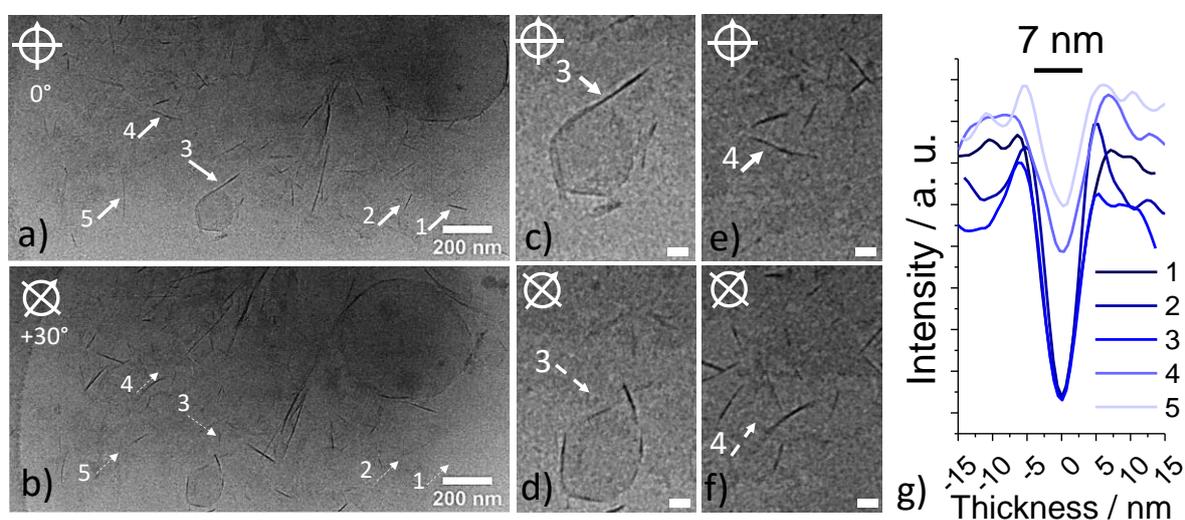

**Figure 3 – Highlight of the bilayer region in cryo-TEM experiments performed on a SL-C18:1-NH$_2$ solution at $C=$ 0.5 wt% and pH 4. Images in c) and e) represent highlights of image in a) corresponding to arrows № 3 and 4. Images in b), d) and f) correspond to the same regions as in a), c) and e) but the sample holder goniometer is tilted of an angle of +30° in order to distinguish the face and sides of bilayers. Segmented arrows in b) point at the same tilted bilayers of the bold arrows in a). g) Cross-section thickness profiles of the bilayers indicated by bold arrows in a). Scale bars in c)-f) correspond to 100 nm.**

In summary, sophorolipids micelles characterized by a neutral amine or a carboxylic acid have an ellipsoidal core ($\frac{R_1}{R}> 1$), surrounded by a prominent hydrophilic region in the equatorial direction ($\frac{T_1}{T}< 1$), as shown in Figure 4a. Upon ionization into ammonium or carboxylate sophorolipids, the repulsive interactions drive the formation of a spheroidal core $\frac{R_1}{R}\sim 1$ surrounded by a prominent hydrophilic region in the polar direction ($\frac{T_1}{T}\geq 1$), as shown in Figure 4b. In the meanwhile, ionization in either ammonium or carboxylate sophorolipids promotes the formation of a biphasic system composed of micelles and flat membrane fragments or platelets (Figure 4b). The latter generally aggregates together and could be detected by bare eye as a suspension in solution. In the ammonium system, the membrane has a thickness consistent with two molecular layers, thus making the hypothesis of a bilayer highly plausible, differently than what we found for other glycolipids, rather assembling in a single interdigitated monolayer.[10,14] The composition and the proportion of the flat objects are still unanswered questions, but it is not excluded that they contain a mixture of neutral and ionic lipids and that their volume fraction is contained below 10% compared to micelles, considering the fact that the SAXS profile is dominated by the micelle signal.

If any explanation trying to associate the nature of the ionized chemical group ($NH_3^+$ vs. $COO^-$) to the morphology (membrane vs. platelet) would be highly speculative at this point, one should consider that unexpected assembling into platelets is not uncommon. Dicephalic imidazole and phosphate surfactants, for instance, are expected to form spherical micelles on the basis of their molecular structure and calculated packing parameter. However, according to their protonation state, they rather form platelets, fibers or even vesicles, instead.[76,77] In particular, flat structures are favored over curved ones despite the presence of charged headgroups expected to promote repulsive interactions. Authors and others attribute such behavior to a complex balance between the close coexistence of ionic and non-ionic headgroups, the presence of directional hydrogen bonds and the screening effects of counterions, shifting repulsive into attractive forces, thus sensibly modifying the effective molecular packing parameter.[76–78]

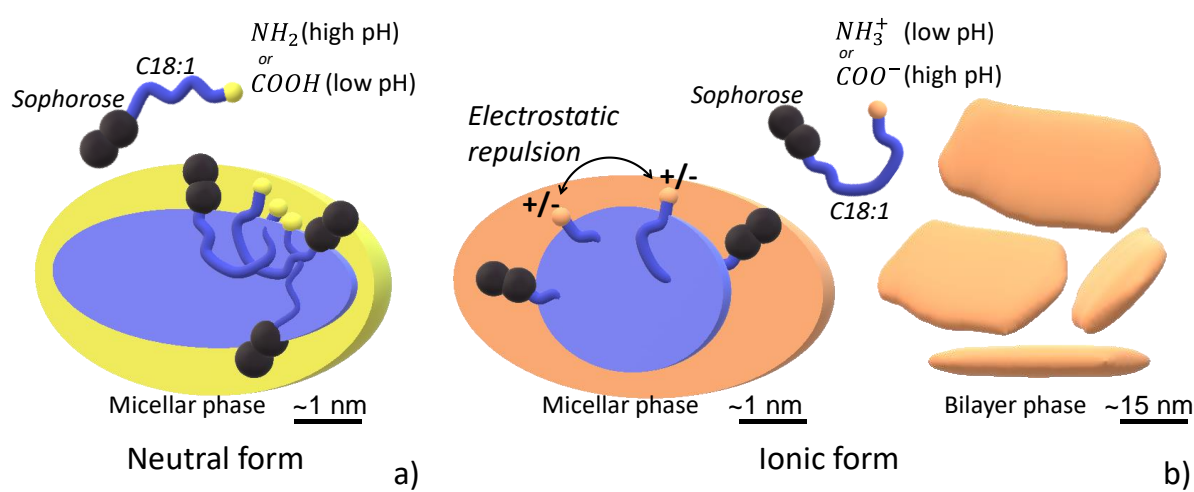

**Figure 4** – **General phase behaviour and micellar structure associated to sophorolipids containing an amine or a carboxylic acid in their a) neutral (–NH₂ and –COOH) and b) ionic ($-NH_3^+$ and $-COO^-$) forms.**

*Saturated aminyl sophorolipids*. The similarities between the phase behavior of sophorolipids containing an amine or a carboxylic acid are also found for the saturated congener, SL-C18:0-NH$_2$. The saturated compound is prepared directly from SL-C18:1-NH$_2$ by catalytic hydrogenation according to a well-established protocol for sophorolipids.[9,10] SL-C18:0-NH$_2$ displays a pH-dependent evolution of both its solubility and colloidal charge: it is unsoluble and neutral at basic pH (-NH$_2$ form, Figure S 6a) while it becomes soluble and positively-charged at acidic pH ($-NH_3^+$ form, Figure S 6a), whereas the transition is settled between 6.5 and 7.5 (empty black circles in Figure S 6a).

The SAXS (Figure 5) study performed on the ionized form of SL-C18:0-NH$_2$ at pH 4.6 shows a typical scattering signal of a micellar system in possible coexistence with larger structures, the former identified by the profile above 0.5 nm$^{-1}$ and the latter by the scattering below 0.5 nm$^{-1}$. The SAXS signal above 0.5 nm$^{-1}$ can be satisfactorily (Figure 5a) fitted with the core-shell model described above and we find $R$= 0.7 nm and $T$= 0.6 nm, with a core aspect ratio of $\frac{R_1}{R}$= 1.5 (Table 2). These values show that ionized SL-C18:0-NH$_2$ micelles have the same size of ionized acidic sophorolipid micelles, as shown by comparing the respective $R$, $T$ and $\frac{R_1}{R}$ in Table 2. The morphology of the objects characterized by the low-$q$ scattering is at the moment unclear, as complementary cryo-TEM experiments could not help answering such question in a satisfactorily way.

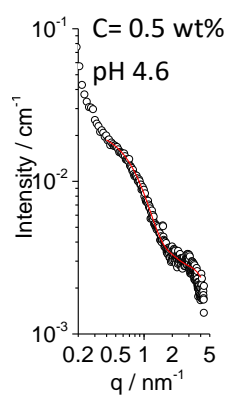

**Figure 5 – SAXS experiment of the SL-C18:0-NH$_2$ solution at pH 4.6 and $C$= 0.5 wt%**

Upon increasing the pH, SL-C18:0-NH$_2$ becomes less charged (Figure S 6a) due to the decrease of amine protonation and solubility decreases. The cryo-TEM study of SL-C18:0-NH$_2$ at pH 8.6 presented in Figure 6b-e shows massive formation of twisted ribbons, with a polydisperse cross section under the present conditions of synthesis. Massive fibrillation is confirmed at a higher scale by SEM-FEG experiments performed on a dried fibrillated SL-C18:0-NH$_2$ sample (Figure 6f), where high magnification (Figure 6g) shows that the aggregates are composed of condensed fibers, indicating that drying and vacuum do not modify the fiber morphology. The corresponding SAXS profile for this sample is shown in Figure 6a (black circles). The main features of the SAXS pattern are the diffraction peak at $q=$ 2.13 nm$^{-1}$, corresponding to a distance of 2.95 nm, and the important low-$q$ scattering having a slope at $q<$ 1 nm$^{-1}$ between -3.5 and -3.7. Similar characteristics were described in the SAXS profiles of saturated acidic sophorolipids, SL-C18:0-COOH, in the same $q$-range,[9] presented on Figure 6a (grey line) as a reference. The flat morphology of the ribbon is expected to provide a -2 slope at low-$q$ region, as reported before.[9] However, the limited window in the present experiments for this specific sample does not allow to reach the -2 slope regime and it only highlights the close to -4 slope, a typical feature of interface scattering. The diffraction peak in self-assembled acidic sophorolipid ribbons was attributed to layered, possibly tilted, crystalline arrangement of the molecules in the plane of the ribbon, on the basis of previous studies on bolaform carboxylic and aminyl glycolipids.[79,80]

The interplanar distance measured in the SL-C18:0-NH$_2$ system is 0.26 nm larger than in the SL-C18:0-COOH system (2.68 nm with $q=$ 2.34 nm$^{-1}$, Figure 6a and Ref. [9]). This value is compatible with the addition of the ethylene diamine moiety grafted onto the COOH group of sophorolipid in SL-C18:0-NH$_2$. The length of an aminoethyl amide chain, if one adds together the C-N$_{amide}$, N-C, C-C and C-N bonds individually is expected to be about 0.58 nm[81] and to this regard, the length of 0.26 nm may seem underestimated. However, one should not forget that the effective lengths of aliphatic molecules are smaller than the direct bond-bond distances; for instance, Tanford uses a value of 0.1265 nm for CH$_2$-CH$_2$ length in aliphatic chains,[82] instead of the bond-bond length of 0.1531 nm.[81] Furthermore, pronounced tilting of the entire SL-C18:0-NH$_2$, resulting in a shorter overall interplanar distance, could be expected, in analogy to similar systems.[80]

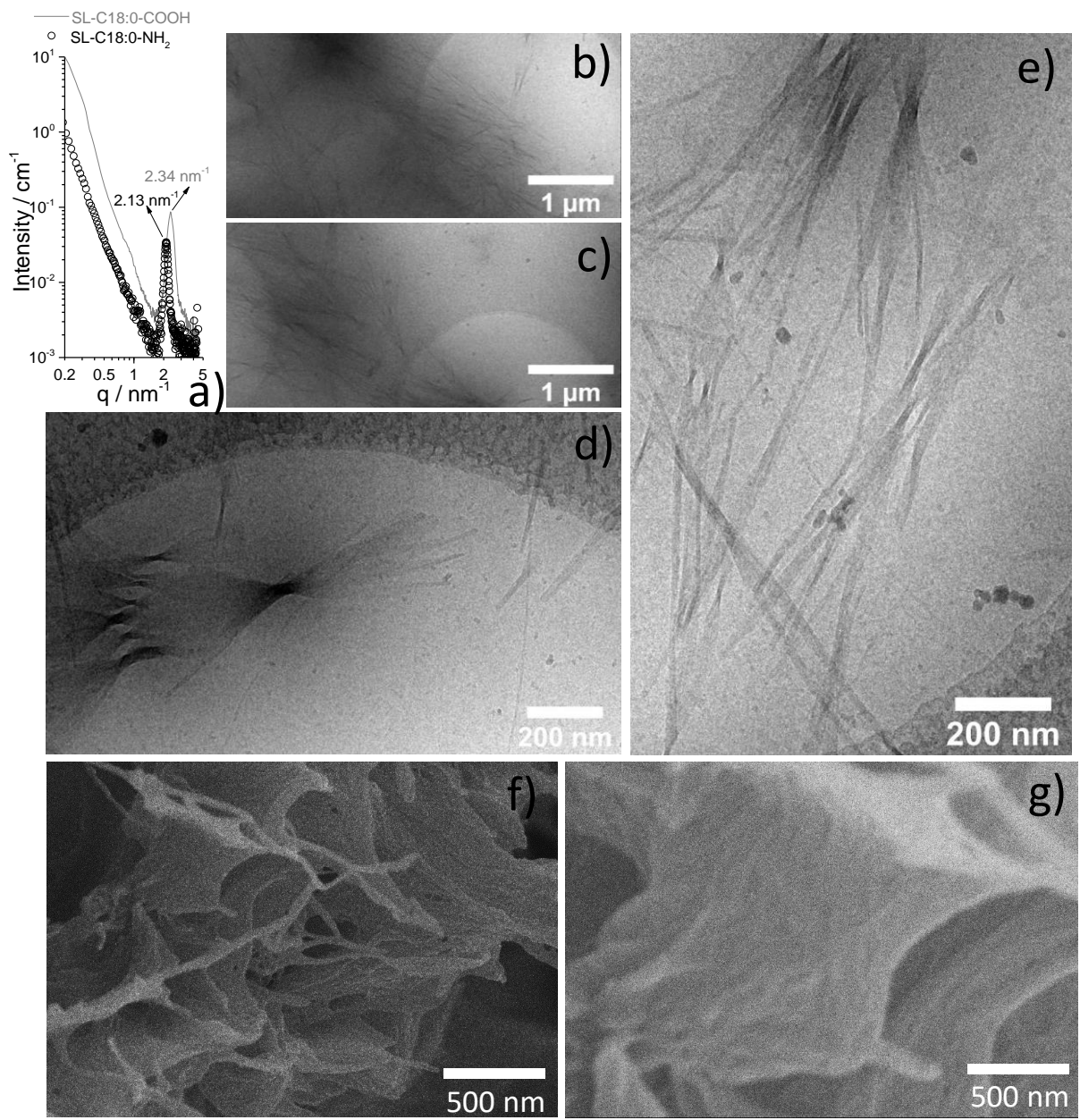

**Figure 6** – a) SAXS profile of a SL-C18:0-NH$_2$ sample recorded at pH 8.5 and $C$= 0.5 wt%. The SAXS profile of twisted ribbons formed by acidic sophorolipids, SL-C18:0-COOH, at equivalent concentration and acidic pH is given in light grey for comparison.[9] b-e) cryo-TEM images of SL-C18:0-NH$_2$ solutions recorded at pH 8.6 and $C$= 0.5 wt%. f-g) SEM-FEG images of dried SL-C18:0-NH$_2$ twisted ribbons prepared at pH 8.6.

In summary, both saturated and monounsaturated sophorolipids show the same self-assembly behavior against ionization and neutrality, independently of the functional end-groups, as summarized in Figure 7. In their neutral form, monounsaturated sophorolipids with an amine or a carboxylic acid self-assemble into ellipsoidal micelles while their saturated derivatives self-assemble into infinitely long twisted ribbons with a sub-50 nm cross section.

In their ionic, ammonium or carboxylate, form, both monounsaturated and saturated sophorolipids self-assemble into spheroidal micelles coexisting with a second minority phase composed of flat aggregated morphologies.

Although surprising, these results are not astonishing, as shown by the similar tendency to assemble into lipid nanotubes for synthetic C18:0 glycolipids bearing a COOH or NH$_2$ group.[79] It then seems that in their neutral forms, bolaform glycolipids with an amine or carboxylic acid group follow the same self-assembly pathway, which generally does not correspond to the structure associated to their expected molecular geometry and packing parameter. For instance, the calculated packing parameter[72] of either SL-C18:0-NH$_2$ or SL-C18:0-COOH falls below 0.3,[10] where spherical micelles are expected, while twisted ribbons are obtained experimentally. Micelles are otherwise experimentally observed in the ionic forms of these compounds, thus respecting the predictions based on the packing parameters. As commented by Svenson,[78] failing of the packing parameter model is not uncommon, especially for surfactants containing functional groups promoting intermolecular hydrogen bonding. This was shown to be the case of bicephalic head-tail imidazole and phosphate surfactants,[76,77] bolaform asymmetrical synthetic carboxylic and aminyl glycolipids[79,80] but also saturated carboxylic and aminyl sophorolipids, discussed in this work. Furthermore, the analogy between carboxylic and aminyl glycolipids, both forming nanotubes,[79,80] and carboxylic and aminyl sophorolipids, both forming either twisted ribbons or micelles respectively according to their charge or neutrality, shows no particular influence of the -NH$_2$ with respect to the -COOH group: both can be charged and undergo intermolecular hydrogen bonding. Failure of the packing parameter approach could also be explained by the different temperature-dependent properties of SL-C18:1-NH$_2$ and SL-C18:0-NH$_2$. DSC experiments of SL-C18:0-NH$_2$ show both a glass-transition temperature, $T_g$= 44.6 ± 0.5°C, and a melting temperature, $T_m$= 107.7°C (Figure S 9a,b), while no particular change in the heat flow is detected for SL-C18:1-NH$_2$ in the range explored at 10°C/min (Figure S 9c).

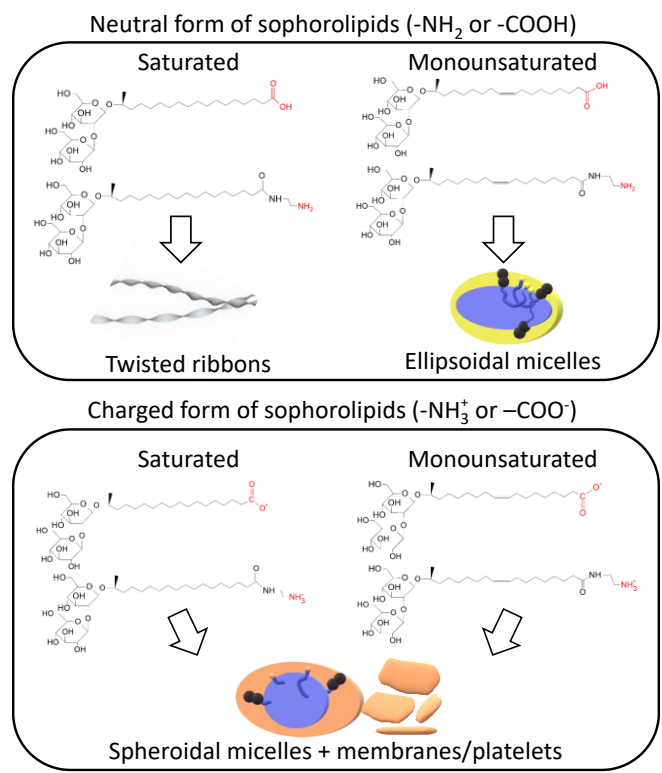

**Figure 7 – Comparison of the self-assembly behavior at room temperature between saturated and monounsaturated carboxylic and aminyl sophorolipids in both the neutral and charged forms.**

*Alkynyl Sophorolipids*. SL-C18:1-C≡CH is obtained by amidation of the COOMe group of C18:1 sophorolipids by reaction with propargylamine. Same for the aminyl derivative, the $^1$H NMR spectrum of SL-C18:1-C≡CH presents all typical signals attributed to sophorolipids (peaks 2, 3, 8, 9, 10, 11, 1´, 1´´ in Figure S 4). The positive and quantitative issue of the reaction is probed by NMR spectroscopy, and in particular by the stoichiometric $^1$H peaks at 2.56 (C≡CH, peak 20 in Figure S 4) and 3.94 ppm (NHCH$_2$C≡CH), peak 21 in Figure S 4) and $^{13}$C peaks at 29.4 (NHCH$_2$C≡CH), 72.0 (C≡CH) and 80.7 (C≡CH), all accounting for the propargylamine group. As for the aminyl derivative presented above, the proof of grafting comes from the signature at 2.19 ppm of the CH$_2$ in position α with respect to C=O (peak 2 in Figure S 4), typical of the amide derivative.[33] SL-C18:1-C≡CH is a white powder, insoluble in water at room temperature. For a possible use of SL-C18:1-C≡CH in aqueous copper click chemistry reactions, it is important to study its phase behavior in water. Considering the replacement of the carboxylic acid, pH is not a possible stimulus to control its solubility, and for this reason, we have rather studied the influence of temperature.

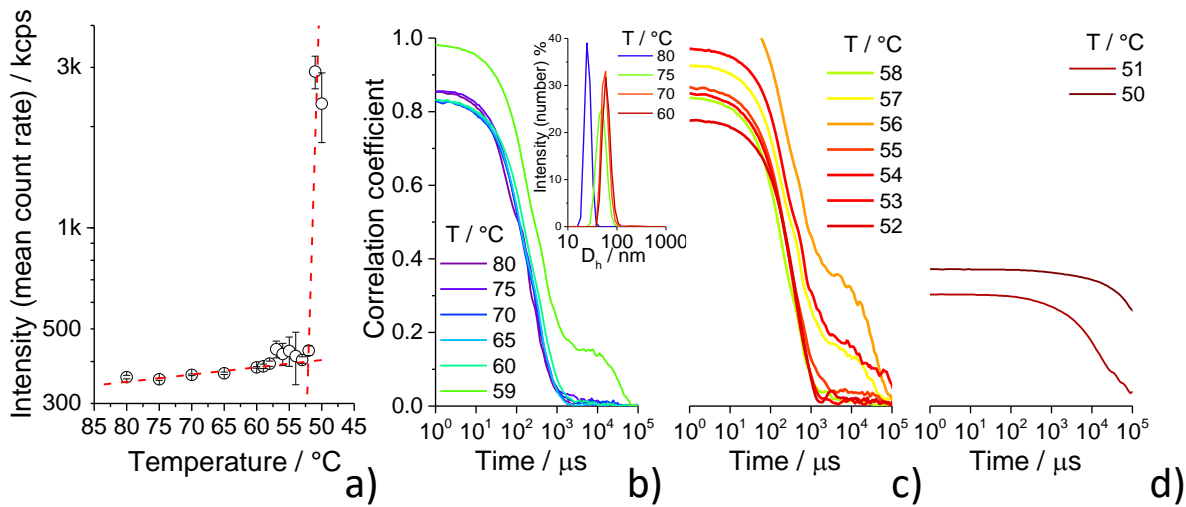

**Figure 8** – Light scattering experiments performed on a SL-C18:1-C≡CH sample solution at $C$= 0.5 wt% as a function of temperature, from 80°C to 50°C: a) static light scattering at 90° and constant shutter (no shutter selected); b-d) dynamic light scattering, correlation coefficient profiles and, in (b), number-weighted intensity distribution of the hydrodynamic diameter, $D_h$.

The temperature of a 0.5 wt% solution of SL-C18:1-C≡CH is first increased to 90°C, at which the solution becomes clear, and then reduced to 25°C, where the compound is not soluble. Light scattering recorded under controlled temperature variation steps is employed to follow the turbidity of the solution between 80°C and 25°C. Figure 8a shows the evolution of the mean count rate recorded at constant shutter when decreasing the temperature. Between 80°C and 60°C, the scattering is quite relevant (400 kcps) although roughly constant and the solution is clear. This may indicate the presence of micellar objects, a hypothesis strengthened by the short plateau (< 10 μs) of the corresponding correlation coefficient and the number size distribution, identifying objects of hydrodynamic diameter, $D_h$, in the order of 20 nm at 80°C and increasing up to 50 nm at 60°C. The SAXS profile of a 0.5 wt% SL-C18:1-C≡CH sample (Figure 9a) recorded on a clear solution prepared at ~90°C (please refer to the methods section for more details concerning the temperature of the sample), is typical of a micellar solution. The corresponding model-independent Guinier plot (Figure 9b) in the $q \cdot R_g <1$ approximation provides $R_g$= 2.52 ± 0.02 and a consequent $R_{Guinier}$= 3.25 ± 0.03. If light scattering and SAXS converge on the fact that the clear solution is constituted by micelles, also confirmed by cryo-TEM experiments showing sub-10 nm micellar objects in the vitrified water layer (Figure 10), they do not precisely converge on the micellar size. At 80°C, the hydrodynamic radius, $R_h$ ($D_h/2$), is about 10 nm, that is about four times $R_{Guinier}$, if one neglects the hydration layer generally contained in the value of $R_h$. We attribute this discrepancy to the impossibility to run SAXS experiments with precise temperature control (please refer to the methods section), to the limited $q$-range of SAXS but also to the approximation of the Guinier model. However, the uncertainty underlying a precise control of temperature does not justify a more detailed model-dependent analysis for this specific system.

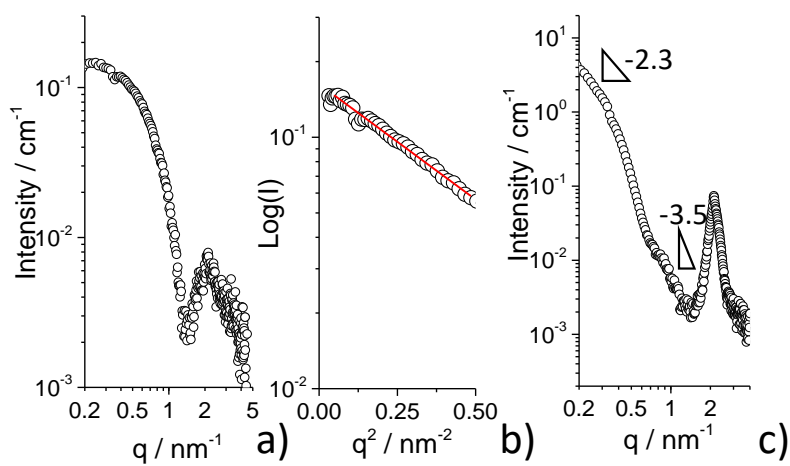

**Figure 9** – SAXS profiles recorded for SL-C18:1-C≡CH solution ($C$= 0.5 wt%) at a,b) ~90°C and c) 25°C after cooling from 90°C. In b), the Guinier plot of the profile in a) in the $qR_g < 1$ region

When the temperature falls below 60°C, the scattering and the error between measurements becomes larger (Figure 8a), suggesting a more heterogeneous medium. This is confirmed by the corresponding correlation coefficient profiles (Figure 8b,c), which show a larger plateau extending up to 100 μs, indicating a growth of the primary micellar phase, but also displaying an additional component above $10^3$ μs, indicating the appearance of objects of much larger size and probably different morphology. Light scattering experiments become less reliable from a quantitative point of view, but they show that the range between 60°C and 52°C certainly constitutes a phase transition region. This is confirmed by the data recorded at 50°C and 51°C, where precipitation occurs macroscopically (Figure 8a) and of which the correlation coefficient profiles (Figure 8d) become unreliable. Linear fitting of the scattering vs. temperature profiles yields a precipitation temperature of 51.8 ± 0.2°C. Complementary DSC experiments performed on the dried powder indicate a $T_g = 48.5 \pm 0.5$ °C, which is in good agreement with the DLS data recorded in solution, and a $T_m = 91.9$ °C, however above the temperature range explored in solution.

Upon temperature decrease to 25°C, the solution presents a massive precipitate, of which the cryo-TEM images (Figure 10b,c) show the formation of twisted ribbons, as previously found for saturated sophorolipids.[9] The ribbon population seems to have a heterogeneous distribution of diameters, a fact which could depend on the cooling protocol employed in this work, whereas cooling rate, uncontrolled in this work, controls homogeneity during crystallization via supersaturation in fibrillar self-assembled low-molecular weight systems.[83] The massive precipitation of twisted ribbons is confirmed by the SEM-FEG experiments in Figure 10d,e and recorded on the dry sample. In this case, drying does not modify the overall sample morphology compared to cryo-TEM and for this reason SEM experiments are representative. The SAXS profile of SL-C18:1-C≡CH at 25°C after heating, shown in Figure 9c, is compatible with the typical profile of twisted ribbons, reported for both the COOH[9] and NH$_2$ (this work) of saturated sophorolipids. The diffraction peak at 2.13 nm$^{-1}$ corresponds to an interplanar distance between SL-C18:1-C≡CH within the ribbon layer of 2.95 nm, of exactly the same length found for the SL-C18:0-NH$_2$ system above, strengthening the hypothesis that longer repeating distance with respect to the ribbons obtained from SL-C18:0-COOH is most likely due to the longer propargylamide end-group. The rest of the SAXS signal is characterized by the expected steep increase in intensity, of slope close to -4 in log-log, typical of interface scattering, but also the beginning of the -2 slope region at the limit of detection, at $q<$ 0.3 nm$^{-1}$.[9,84,85]

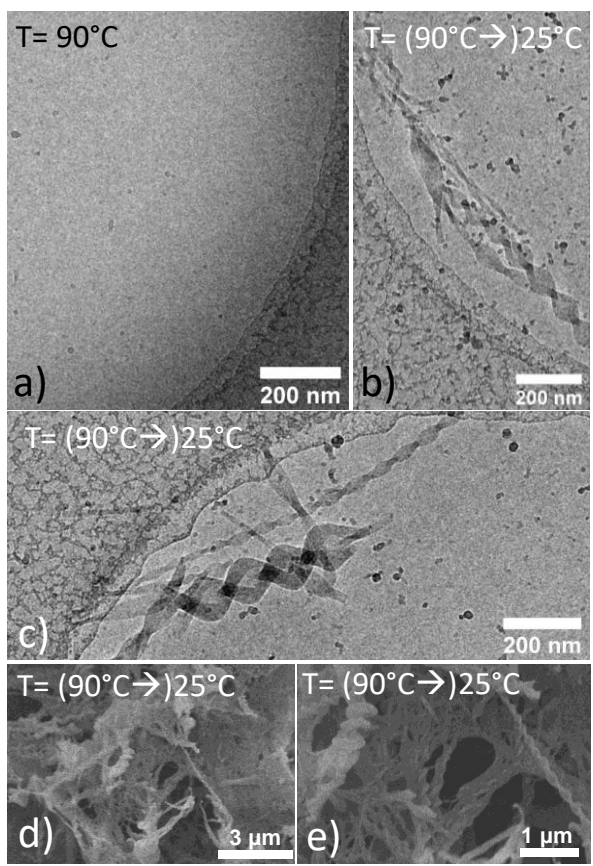

**Figure 10** – a-c) Cryo-TEM and d,e) SEM-FEG images recorded on a SL-C18:1-C≡CH sample ($C$= 0.5 wt%) at (a) 90°C and (b-e) 25°C after cooling from 90°C

**Conclusion**

Deacetylated aminyl (SL-C18:1-NH$_2$) and alkynyl (SL-C18:1-C≡CH) derivatives of monounsaturated (C18:1) sophorolipids are obtained by amidation reaction between a methyl ester derivative of sophorolipids and, respectively, ethylenediamine and propargylamine. The saturated aminyl derivative (SL-C18:0-NH$_2$) is obtained by catalytic hydrogenation of the corresponding monounsaturated molecule. The aminyl derivative is prepared in methanol at 120°C while the alkynyl derivative is prepared in THF at 50°C by enzymatic catalysis. The self-assembly in water under dilute conditions is studied for all compounds by means of light scattering, SAXS, SANS and cryo-TEM.

SL-C18:1-NH$_2$ forms a micellar phase in the pH range between about 4 and 11. In its neutral $NH_2$ state at high pH, micelles are ellipsoids of revolution with the equatorial and polar radii being, respectively, $R_{CB\ model}$= 1.75 ± 0.18 nm and $R_{1\ CB\ model}$= 3.21 ± 0.32 nm, estimated with the "*coffee bean*" micellar model. The radius estimated using the model-independent Guinier approach is $R_{Guinier}$= 2.44 ± 0.05 nm, in good agreement with the model. In its ionized $NH_3^+$ state at low pH, confirmed by electrophoretic mobility experiments, $R_{Guinier}$= 1.99 ± 0.08 nm, while $R_{CB\ model}$= 1.75 ± 0.18 nm and $R_{1\ CB\ model}$= 2.48 ± 0.25 nm. If $R_{CB\ model}$ is practically unchanged under basic and acidic conditions, the "*coffee bean*" micellar model shows a considerable difference in the core and shell aspect ratios between the neutral and ionized states. Neutral sophorolipid micelles are ellipsoids with an elongated core and a thick equatorial shell, while ionized micelles have a spheroidal core with a thick polar shell. This behavior is in complete agreement with what was found for acidic sophorolipid (SL-C18:1-COOH) micelles, the only obvious difference being the inversed ionization pH for the latter (basic) with respect to the aminyl derivative (acidic). Also in agreement with the self-assembly behavior of ionized ($COO^-$) acidic sophorolipids, ionized ($NH_3^+$) SL-C18:1-NH$_2$ solutions at acidic pH display the coexistence between the micellar phase and membrane phase, constitute of bilayer fragments having a thickness of about 7 nm.

The analogy between COOH and $NH_2$ sophorolipids also exists for saturated SL-C18:0-NH$_2$ sophorolipids. In its low-pH ionized ($NH_3^+$) state, shown by electrophoretic mobility experiments, this compound forms a micellar phase, where micelles are spheroids of equatorial radius $R_{CB\ model}$= 1.30 ± 0.13 nm. Upon neutralization of the positive charge when increasing pH, SL-C18:0-NH$_2$ undergoes a micelle-to-twisted ribbon phase transition. The ribbons are semi-crystalline soft solids with an interplanar distance of 2.95 nm between sophorolipid layers within the ribbon plane. This value is less than 0.3 nm larger for the interplanar distance fund in SL-C18:0-COOH twisted ribbons, thus suggesting a similar molecular packing. We estimate that the increased distance can be attributed to the ethylene diamine molecule grafted at the tip of sophorolipids.

Alkynyl sophorolipids SL-C18:1-C≡CH do not have pH but rather temperature-responsive properties. Insoluble at room temperature, solubility in water is achieved above about 52°C, a temperature close to its $T_g$= 48.5 ± 0.5 °C. We have detected a major micellar phase at about 90°C with $R_{Guinier}$= 3.25 ± 0.03 nm and a twisted ribbon phase with a repeating interlipid distance of 2.95 nm below the transition temperature.

This work brings an additional brick to the chemical derivatization of sophorolipid biosurfactants. We show that aminyl and alkynyl sophorolipids can be easily prepared, thus paving the way to further derivatization for a new family of functional sophorolipids. At the same time, from a fundamental point of view, we also show that replacing the COOH by the $NH_2$ group reverses the charge and responsivity to pH but it does not sensibly modify the self-assembly behavior, neither for the C18:1 nor for the C18:0 derivatives.

**Acknowledgements**

The authors acknowledge IMPC (Institut des Matériaux de Paris Centre, FR2482) and the C'Nano projects of the Region Ile-de-France, for SEM-FEG apparatus funding and David Montero for his assistance in the SEM-FEG experiments. SANS experiments were performed at the D11 beamline of the Institute Laue-Langevin (ILL) during the proposal number 9-13-778. We kindly acknowledge Dr. Andrea Lassenberger and Sylvain Prévost for their kind assistance during the experiments. We thank ILL for financial support of the beamtime. SAXS experiments were performed at the European Radiation Synchrotron Facility (ESRF) during the proposal number SC-4639. Dr. Daniel Hermida-Merino is kindly acknowledged for his assistance during the experiments. We also acknowledge ESRF for financial support of the beamtime. This work was supported by a grant overseen by the French National Research Agency (ANR) (SELFAMPHI-19-CE43-0012-01).

# Synthesis and self-assembly of aminyl and alkynyl substituted sophorolipids

Abdoul Aziz Ba,[a] Jonas Everaert,[b] Alexandre Poirier,[a] Patrick Le Griel,[a] Wim Soetaert,[c] Sophie L. K. W. Roelants,[c] Daniel Hermida-Merino,[d] Christian V. Stevens,[b] Niki Baccile[a,*]

[a] Sorbonne Université, Centre National de la Recherche Scientifique, Laboratoire de Chimie de la Matière Condensée de Paris, LCMCP, F-75005 Paris, France

[b] SynBioC, Department of Green Chemistry and Technology, Ghent University, Ghent, Belgium

[c] InBio, Department of Biotechnology, Ghent University, Ghent, Belgium

[d] Netherlands Organisation for Scientific Research (NWO), DUBBLE@ESRF BP CS40220, 38043 Grenoble, France

*Synthesis procedure, $^1$H and $^{13}$C NMR analyses of modified sophorolipids*

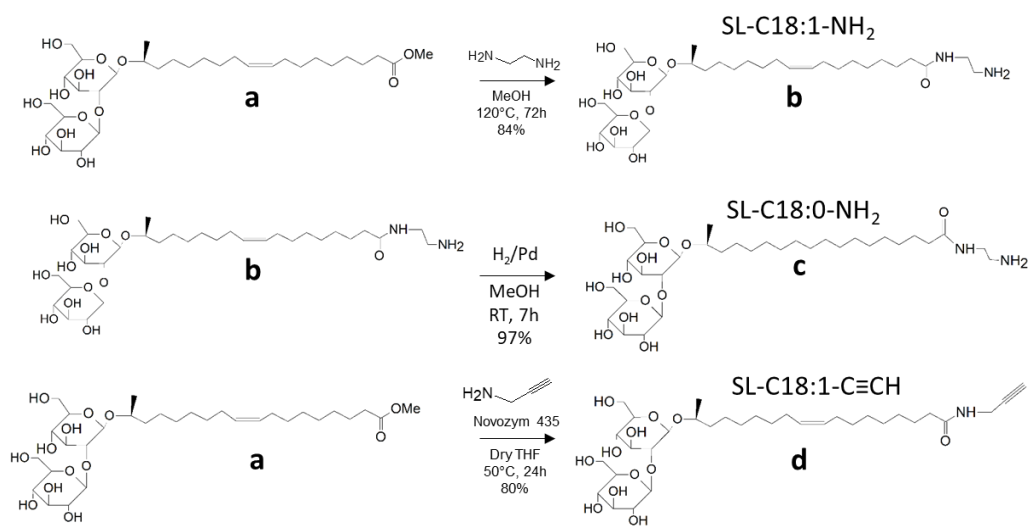

**Figure S 1** – Synthesis conditions for aminyl and alkynyl sophorolipids. (a) Monounsaturated sophorolipids methyl ester, SL-C18:1-OMe, prepared from a previous study.[1] (b) Monounsaturated aminyl sophorolipid, SL-C18:1-NH$_2$. (c) Saturated aminyl sophorolipid, SL-C18:0-NH$_2$. (d) Monounsaturated alkynyl sophorolipid, SL-C18:1-C≡CH.

*Monounsaturated sophorolipid methyl ester, SL-C18:1-OMe* (Figure S 1a). *SL-C18:1-OMe* is prepared according to a previous study[1] from lactonic sophorolipids (LSL)[1] and following a literature process[2,3] In a 100 mL round-bottom flask, sodium methylate is formed *in situ* by adding 0.083 g of sodium (3.63 mmol, 0.5 eq) to 20 mL of anhydrous MeOH and 5 g of LSL (7.26 mmol). The flask is equipped with a reflux condenser and a tube containing CaCl$_2$ to protect the reaction mixture from atmospheric humidity. The reaction mixture is stirred for 3 hours at reflux temperature, cooled to room temperature and acidified to neutral pH with acetic acid. The mixture was concentrated under reduced pressure, dissolved in deionized water and cooled to 0°C in an ice bath. The sophorolipid methyl ester (**b**) precipitates as a white powder. The precipitate is filtered, washed with water and dried under reduced pressure (79%, 3.65 g). The molecule in Figure S 1**a** is identified by $^1$H and $^{13}$C NMR (Figure S 2). The allocation of the peaks is in agreement with the data relating to this compound.[3]

*Monounsaturated aminyl sophorolipid, SL-C18:1-NH$_2$* (Figure S 1**b**): in a 50 mL pressurized oven-dried bottle, 2 g of (**a**) (3.14 mmol, 1 eq) is dissolved in 12 mL of anhydrous MeOH, to which 2.12 ml of ethylenediamine (31.4 mmol, 10 equivalents, 1.89 g) is added. The reaction mixture is heated at 120°C for 72 hours with magnetic stirring. The reaction is followed by NMR analysis until complete conversion. The solvent of the reaction mixture is evaporated *in vacuo* and the product is purified by chromatography on silica gel (10% H$_2$O, 25% MeOH, 15% Et$_3$N, 50% EtOAc). The desired product is obtained in the form of a viscous brown oil (84%, 1.75 g) and identified by $^1$H and $^{13}$C NMR. Attributions of the $^1$H and $^{13}$C NMR signals are given below. The corresponding $^1$H NMR spectrum is given in Figure S 3.

**$^1$H NMR (400 MHz, MeOD-d4):** δ = 1.25 (3H, d, *J*=6.2 Hz, C<u>H</u>$_3$CH), 1.29-1.48 (17H, m, C<u>H</u>$_a$H$_b$CHCH$_3$, 8xC<u>H</u>$_2$(CH$_2$)$_2$), 1.58-1.65 (3H, m, CH$_a$<u>H</u>$_b$CHCH$_3$, C<u>H</u>$_2$CH$_2$CONH), 2.02-2.08 (4H, m, 2xC<u>H</u>$_2$CH=CH), 2.19 (2H, t, *J*=7.5 Hz, C<u>H</u>$_2$CONH), 2.89 (2H, t, *J*=6.1 Hz, C<u>H</u>$_2$NH$_2$), 3.21-3.41 (8H, m, 6xC<u>H</u>OC, CONHC<u>H</u>$_2$), 3.45 (1H, dxd, *J*=8.4 Hz, *J*=8.4 Hz, C<u>H</u>OC), 3.56 (1H, dxd, *J*=8.7 Hz, *J*=8.7 Hz, C<u>H</u>OC), 3.63-3.69 (2H, m, 2xC<u>H</u>$_a$H$_b$OH), 3.79-3.89 (3H, m, 2xCH$_a$<u>H</u>$_b$OH, C<u>H</u>CH$_3$), 4.45 (1H, d, *J*=7.7 Hz, C<u>H</u>(O)$_2$), 4.64 (1H, d, *J*=7.8 Hz, C<u>H</u>(O)$_2$), 5.34- 5.38 (2H, m, C<u>H</u>=C<u>H</u>). **$^{13}$C NMR (100.6 MHz, MeOD-d4):** δ = 21.9 (<u>C</u>H$_3$CH), 26.2 (<u>C</u>H$_2$(CH$_2$)$_2$), 26.8 (<u>C</u>H$_2$CH$_2$CONH), 28.1 (2x<u>C</u>H$_2$CH=CH), 30.2 (<u>C</u>H$_2$(CH$_2$)$_2$), 30.3 (2x<u>C</u>H$_2$(CH$_2$)$_2$), 30.4 (<u>C</u>H$_2$(CH$_2$)$_2$), 30.8 (2x<u>C</u>H$_2$(CH$_2$)$_2$), 30.9 (<u>C</u>H$_2$(CH$_2$)$_2$), 37.0 (<u>C</u>H$_2$CONH), 37.8 (<u>C</u>H$_2$CHCH$_3$), 40.4 (CONH<u>C</u>H$_2$), 41.4 (<u>C</u>H$_2$NH$_2$), 62.7 (<u>C</u>H$_2$OH), 63.1 (<u>C</u>H$_2$OH), 71.5 (<u>C</u>HOC), 71.8 (<u>C</u>HOC), 75.9 (<u>C</u>HOC), 77.8 (2x<u>C</u>HOC), 78.2 (<u>C</u>HOC), 78.3 (<u>C</u>HOC), 78.9 (<u>C</u>HCH$_3$), 81.9 (<u>C</u>HOC), 102.7 (<u>C</u>H(O)$_2$), 104.7 (<u>C</u>H(O)$_2$), 130.8 (CH=<u>C</u>H), 130.9 (<u>C</u>H=CH), 177.2 (<u>C</u>ONH).

*Saturated aminyl sophorolipid*, SL-C18:0-NH$_2$ (Figure S 1**c**): 1.75 g of **b** is dissolved in 30 mL of MeOH under an argon atmosphere and to which 175 mg (10% w/w) of Pd/C (10%) is added. The reaction mixture is stirred for 7 hours under an atmosphere of 5 bars of H$_2$, after which it is filtered through celite. After removing the solvent under vacuum, a white solid of saturated **aminyl** sophorolipid is obtained (1.70 g, 97% yield), as identified by the loss of the C<u>H</u>=C<u>H</u> peak at 5.37 ppm in solution $^1$H NMR, as reported for a similar reaction on acidic sophorolipids.[4,5]

*Monounsaturated alkynyl sophorolipid*, SL-C18:1-C≡CH (Figure S 1**d**): in a dried 50 mL flask, 1.8 g of (**a**) (3.14 mmol, 1 eq) is dissolved in 30 mL of anhydrous THF. 0.6 g of Novozym 435 (33% by weight of sophorolipid) and 361 μL of propargylamine (5.65 mmol, 2 equivalents, 0.311 g) are added and the reaction mixture is heated at 50°C for 24 hours under magnetic stirring. The reaction is followed by NMR until complete conversion. The reaction mixture is filtered through a sintered glass filter and the solvent is evaporated *in vacuo*. The product is purified by chromatography on silica gel (5% H$_2$O, 20% MeOH, 75% EtOAc). The desired product is obtained in the form of a yellowish powder (80%, 1.5 g). Attributions of the

$^{1}$H and $^{13}$C NMR signals are given below. The corresponding $^{1}$H NMR spectrum is given in Figure S 4.

**$^{1}$H NMR (400 MHz, MeOD-d4):** δ = 1.25 (3H, d, *J*=6.3 Hz, C<u>H</u>$_3$CH), 1.30-1.49 (17H, m, CH$_a$<u>H</u>$_b$CHCH$_3$, 8xC<u>H</u>$_2$(CH$_2$)$_2$), 1.56-1.65 (3H, m, C<u>H</u>$_a$H$_b$CHCH$_3$, C<u>H</u>$_2$CH$_2$CONH), 2.03-2.04 (4H, m, 2xC<u>H</u>$_2$CH=CH), 2.19 (2H, t, *J*=7.5 Hz, C<u>H</u>$_2$CONH), 2.56 (1H, t, *J*=2.5 Hz, C≡C<u>H</u>), 3.21-3.33 (5H, m, 5xC<u>H</u>OC), 3.37 (1H, dxd, *J*=8.9 Hz, *J*=8.9 Hz, C<u>H</u>OC), 3.45 (1H, m, C<u>H</u>OC), 3.55 (1H, dxd, *J*=8.7 Hz, *J*=8.7 Hz, C<u>H</u>OC), 3.63-3.68 (2H, m, 2xC<u>H</u>$_a$H$_b$OH), 3.79-3.88 (3H, m, 2xCH$_a$<u>H</u>$_b$OH, C<u>H</u>CH$_3$), 3.94 (2H, d, *J*=2.5 Hz, NHC<u>H</u>$_2$C≡C), 4.45 (1H, d, *J*=7.7 Hz, C<u>H</u>(O)$_2$), 4.64 (1H, d, *J*=7.8 Hz, C<u>H</u>(O)$_2$), 5.34-5.37 (2H, m, C<u>H</u>=C<u>H</u>). **$^{13}$C NMR (100.6 MHz, MeOD-d4):** δ = 21.9 (<u>C</u>H$_3$CH), 26.2 (<u>C</u>H2(CH$_2$)$_2$), 26.9 (<u>C</u>H$_2$CH$_2$CONH), 28.1 (<u>C</u>H$_2$CH=CH), 28.2 (<u>C</u>H$_2$CH=CH), 29.4 (NH<u>C</u>H$_2$C≡C), 30.2 (2x<u>C</u>H$_2$(CH$_2$)$_2$), 30.3 (<u>C</u>H$_2$(CH$_2$)$_2$), 30.4 (<u>C</u>H$_2$(CH$_2$)$_2$), 30.8 (<u>C</u>H$_2$(CH$_2$)$_2$), 30.8 (<u>C</u>H$_2$(CH$_2$)$_2$), 30.9 (<u>C</u>H$_2$(CH$_2$)$_2$), 36.8 (<u>C</u>H$_2$CONH), 37.8 (<u>C</u>H$_2$CHCH$_3$), 62.8 (<u>C</u>H$_2$OH), 63.1 (<u>C</u>H$_2$OH), 71.5 (<u>C</u>HOC), 71.8 (<u>C</u>HOC), 72.0 (C≡<u>C</u>H), 75.9 (<u>C</u>HOC), 77.8 (2x<u>C</u>HOC), 78.2 (<u>C</u>HOC), 78.3 (<u>C</u>HOC), 78.9 (<u>C</u>HCH$_3$), 80.7 (<u>C</u>≡CH), 81.9 (<u>C</u>HOC), 102.7 (<u>C</u>H(O)$_2$), 104.7 (<u>C</u>H(O)$_2$), 130.8 (CH=<u>C</u>H), 130.9 (<u>C</u>H=CH), 175.9 (<u>C</u>ONH).

*Solution Nuclear Magnetic Resonance* (*NMR*): $^{1}$H NMR and $^{13}$C NMR spectra were recorded at 25 °C at 400 MHz and 100.6 MHz, respectively using a Bruker AVANCE III HD 400 Nanobay spectrometer, equipped with 1H/BB z-gradient probe (BBO, 5 mm). Chemical shifts (δ) are reported in parts per million (ppm) relative to tetramethylsilane (δ = 0) and referenced to the residual solvent peak (MeOD-d4 δ$_H$ = 3.31 and δ$_C$ = 49.0). All spectra were processed using TOPSPIN 3.2. $^{1}$H, $^{13}$C (APT), COSY, HSQC and HMBC NMR spectra were acquired through the standard sequences available in the Bruker pulse program library.

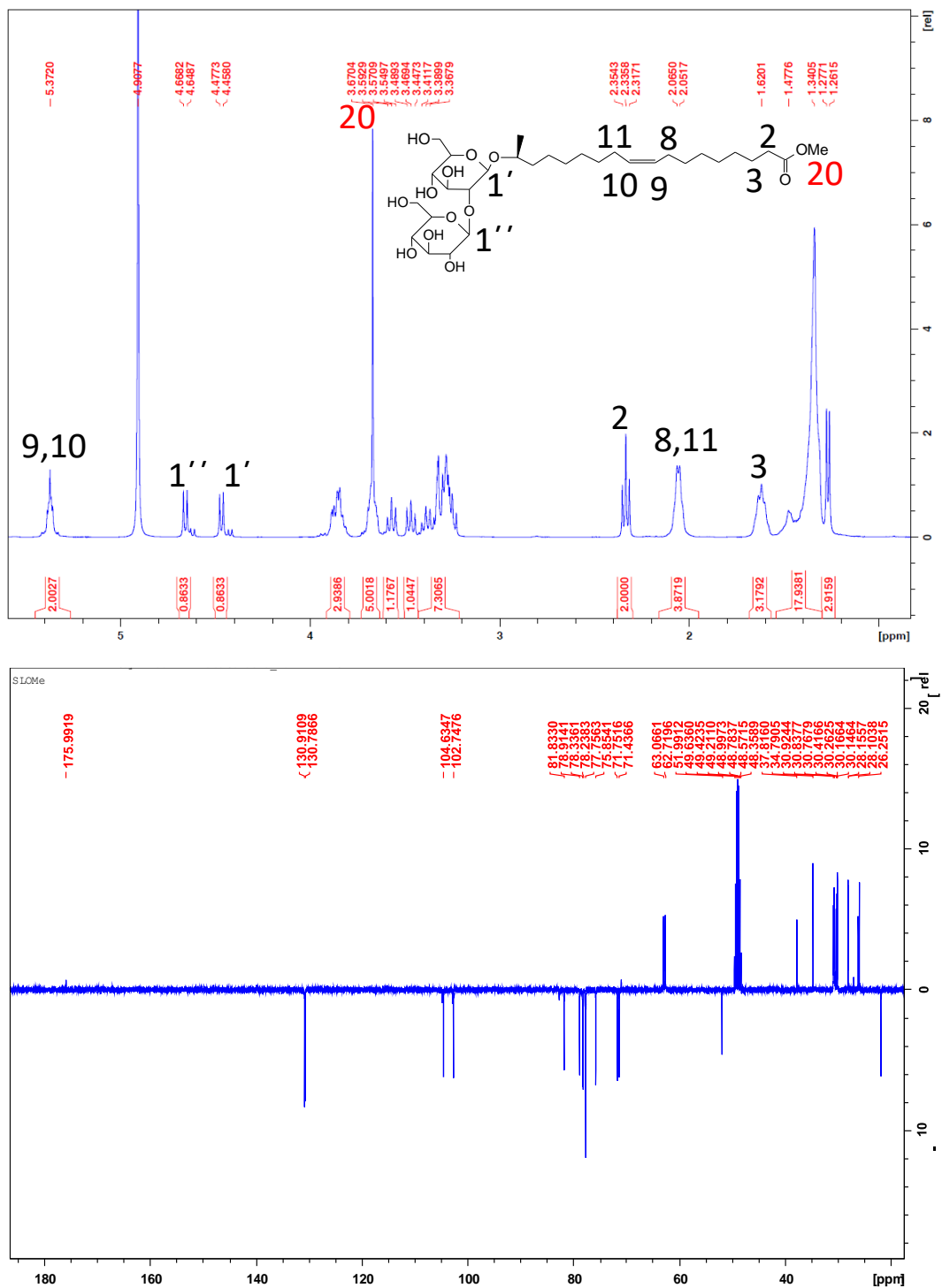

**Figure S 2** – $^1$H and $^{13}$C (APT) solution NMR spectrum in MeOD-d4 of monounsaturated aminyl sophorolipid, SL-C18:1-OMe

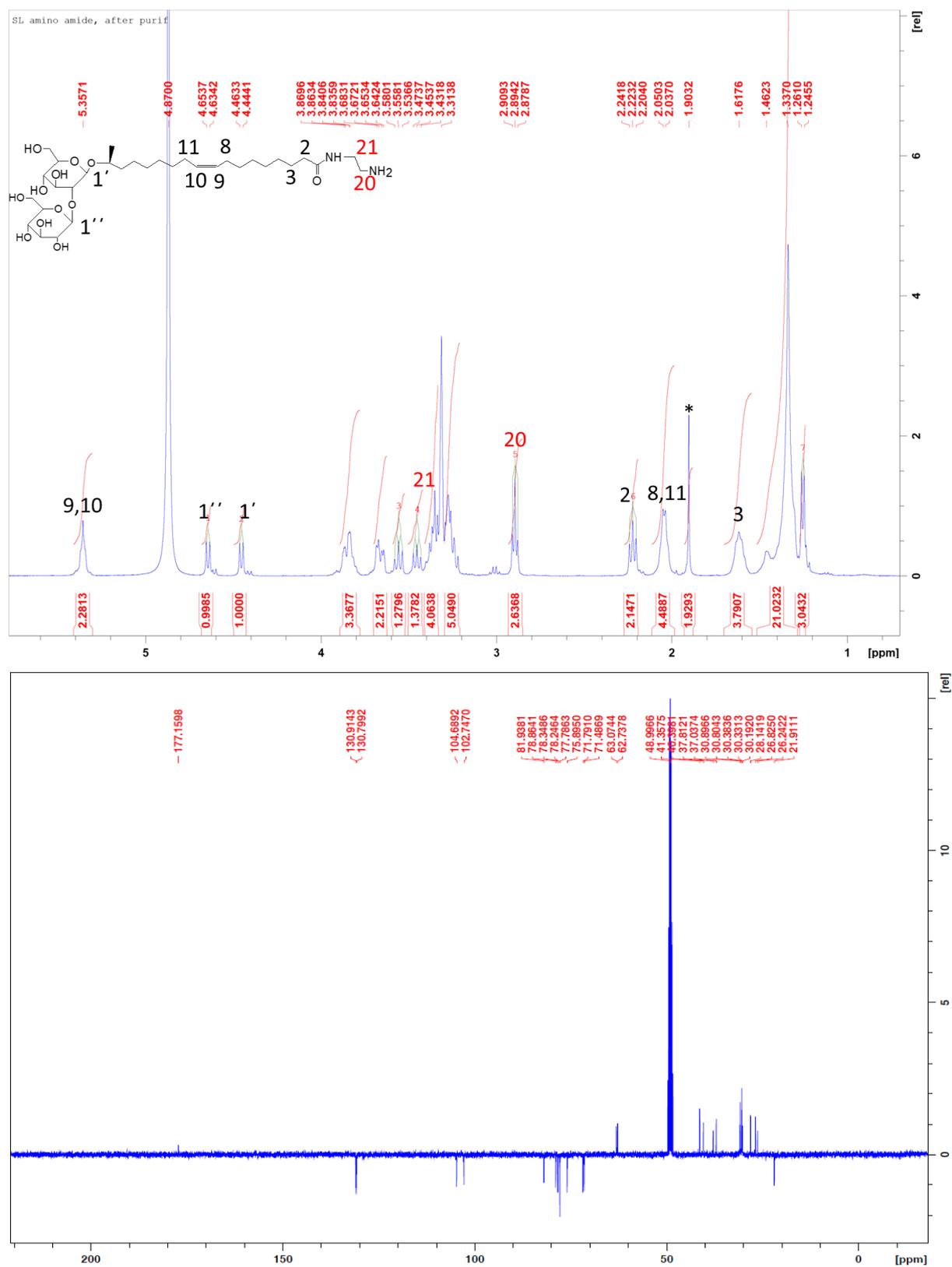

**Figure S 3** – $^1H$ and $^{13}C$ (APT) solution NMR spectrum in MeOD-d4 of monounsaturated aminyl sophorolipid, SL-C18:1-NH$_2$. The * symbol in the $^1H$ NMR spectrum indicates a contaminant in the NMR tube.

**Figure S 4** - ¹H and ¹³C (APT) solution NMR spectrum in MeOD-d4 of sophorolipid alkynyl, SL-C18:1-C≡CH

*Small Angle X-ray Scattering (SAXS).* SAXS experiments are performed at the DUBBLE BM26B beamline at the ESRF synchrotron facility (Grenoble, France).[6,7] Samples have been analyzed during the run SC4639 using a beam at 11.93 KeV and a sample-to-detector distance of 2.10 m. Samples are analyzed in quartz capillaries of 2 mm in diameter, including a water solution and an empty capillary, respectively used to subtract the background signal and measure the scattering level of water ($I(0)$= 0.016 cm$^{-1}$) for absolute scale calibration. The signal of the Pilatus 1M 2D detector (172 x 172 µm pixel size), used to record the data, is integrated azimuthally with PyFAI to obtain the $I(q)$ spectrum ($q = 4\pi \sin\theta / \lambda$, where $2\theta$ is the scattering angle) after masking systematically wrong pixels and the beam stop shadow. Silver behenate ($d_{(100)}$ = 58.38 Å) is used as SAXS standard to calibrate the $q$-scale. Experiments are performed at room temperature (23 ± 2°C) for SL-C18:1-NH$_2$ and SL-C18:0-NH$_2$ solutions. The SL-C18:1-C≡CH solution was on the contrary measure both at room temperature and after heating to 90°C. However, considering the fact that the beamline did not dispose of a temperature-controlling unit during beamtime, the 90°C could not be carefully controlled. In practice, the capillary containing the SL-C18:1-C≡CH solution is heated at 90°C for about 5 min, when the solution becomes clear. The capillary is then transferred as fast as possible in front of the beam and the experiment run immediately after. All in all, between removal from the 90°C source and the acquisition we estimate about 2 to 3 min, which could cause a loss in effective temperature of the solution in the capillary of few degrees. Although we cannot guarantee that acquisition occurred when the solution was at 90°C, the solution was still clear during the analysis and no precipitation occurred. To avoid confusion on this point and to indicate that the temperature is not strictly controlled, we employ the notation T= ~90°C when referring to the SAXS experiment in the text.

*Small Angle Neutron Scattering (SANS).* SANS experiments have been performed at the D11 beamline of Institut Laue Langevin (Grenoble, France). Four $q$-ranges have been explored and merged using the following wavelengths, λ, and sample-to-detector (StD) distances. 1) ultra-low-$q$: λ= 13.5Å, StD= 39 m; 2) low-$q$: λ= 5.3Å, StD= 39 m; 3) mid-$q$: λ= 5.3Å, StD= 8 m; 4) high-$q$: λ= 5.3Å, StD= 1.4 m. All samples are prepared in 99.9% D$_2$O (including the use of using NaOD and DCl solutions for pH change) to limit the incoherent background scattering. Solutions are analyzed in standard 1 mm quartz cells. Direct beam, empty cell, H$_2$O are recorded and boron carbide (B$_4$C) is used as neutron absorber. The background sample (D$_2$O) signal was subtracted from the experimental data. Absolute values of the scattering intensity are obtained from the direct determination of the number of neutrons in the incident beam and the detector cell solid angle. The 2D raw data were corrected for the ambient background and empty cell scattering and normalized to yield an absolute scale (cross section per unit volume) by the neutron flux on the samples. The data were then circularly averaged to yield the 1D intensity distribution, $I(q)$. The software package Grasp (developed at ILL and available free of charge) is used to integrate the data, while the software package SAXSUtilities (developed at ESRF and available free of charge) is used to merge the data acquired at all configurations and subtract the background. Experiments are thermalized at 25°C using the beamline sample temperature controller.

*Analysis of the scattering data.* SAXS and SANS data were analyzed using a model-dependent and model-independent approach. The low-$q$ region below $q<$ 0.2 nm$^{-1}$ is analyzed using a classical evaluation of the slope of $I(q)$ in a log-log scale. The data are fitted using a linear function, of which the slope is generally related to a specific morphology (e.g., -1: cylinders; -2: lamellae),[8] or it describes the presence of fractal objects.[9] The region between ~0.2 < $q$ / nm$^{-1}$ < ~5 is analyzed with both model-independent (Guinier) and model-dependent (core-shell prolate ellipsoid of revolution form factor) functions.

The model-independent Guinier analysis can be safely applied to those data showing a plateau below $q<$ 0.2 nm$^{-1}$.[8] Within the Guinier approximation $q.R_g<$ 1, with $q$ being the wavevector and $R_g$ the radius of gyration, the scattered intensity can be approximated by

$$I(q) = I_0 e^{-\frac{R_g^2 q^2}{3}}$$

which, expressed in the log-log plot, gives

$$Log(I) = Log(I_0) - \frac{Log(e)R_g^2}{3}q^2 = a + bq^2.$$

$R_g$ is obtained by linearization and plotting $Log(I)$ against $q^2$, with the slope being equal to $b = -0.434\frac{R_g^2}{3}$, with $Log(e) = 0.434.$ For a spherical object, one can estimate the radius of the corresponding sphere, $R$, according to

$$R_{Guinier} = \sqrt{\frac{5}{3}R_g^2} = \sqrt{5\cdot(-\frac{b}{0.434})}$$

The error on $R_{Guinier}$ is derived from the error of the linear fit.

The model-dependent analysis consists in employing a core-shell (prolate) ellipsoid of revolution form factor model with inhomogeneous shell thickness, also referred to as the "*coffee-bean*" model, in agreement with previous modelling of SAXS data recorded on deacetylated acidic C18:1 sophorolipids.[10,11] The model is implemented in the SasView 3.1.2 software (CoreShellEllipsoidXT),[a] the general equation of which is

$$I(q) = \frac{scale}{V}(\rho - \rho_{solv})^2 P(q) S(q) + bkg$$

where, $scale$ is the volume fraction, $V$ is the volume of the scatterer, $\rho$ is the Scattering Length Density (SLD) of the object, $\rho_{solv}$ is the SLD of the solvent, $P(q)$ is the form factor of the object, $bkg$ is a constant accounting for the background level and $S(q)$ is the structure factor, which is hypothesized as unity in the analyzed range of $q$-values and at the present concentrations. The analytical expression of $P(q)$ for a core-shell ellipsoid of revolution model implemented in the software is provided on the developer's website[a], while Figure S 5 shows the geometrical model, where $T$ is the equatorial shell thickness, $T_1$ is the polar shell thickness, $R$, the equatorial core radius, $R_1$, the polar core radius. The model implies the evaluation of $\rho_{core}$, $\rho_{shell}$, $\rho_{solv}$, the SLDs of, respectively, the hydrophobic core, hydrophilic shell and solvent. The model also considers a non-homogeneous core and shell, for we define the aspect ratio of the core and shell respectively being $\frac{R_1}{R}$ and $\frac{T_1}{T}$. The SLD can be calculated using the SLD calculator implemented in the SasView 3.1.2 software and based on

$$\rho = \frac{\sum_i^j Z_i\, r_e}{v_M}$$

where $Z_i$ is the atomic number of the $i^{th}$ of $j$ atoms in a molecule of molecular volume $v_M$, $r_e$ is the classical electron radius or Thomson scattering length (2.8179 × 10$^{-15}$ m). The list of fixed and variable in our approach is given in Figure S 5

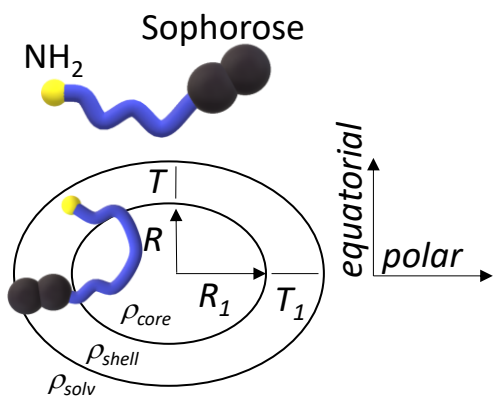

| Parameter | Value |
|---|---|
| $T$ | Variable |
| $T_1/T$ | Variable |

---

[a] http://www.sasview.org/sasview/user/models/model_functions.html#coreshellellipsoidxtmodel

| $R$ | Variable |
|---|---|
| $R_1/R$ | Variable |
| $\rho_{shell}$ | Variable |
| $\rho_{core}$ | $8.4 \times 10^{-4}$ nm$^{-2}$ |
| $\rho_{solv}$ | $9.4 \times 10^{-4}$ nm$^{-2}$ |

**Figure S 5 – The "*coffee-bean*" micellar model: core-shell prolate ellipsoid of revolution form factor model with inhomogeneous shell thickness ($T_1 \neq T$) adapted from Ref. [10,11] and available in the SasView 3.1.2 software (CoreShellEllipsoidXTModel).[a] List of the main fixed and variable parameters used in the model to fit the SAXS data.**

The values of 8.4 and 9.4 x 10$^{-4}$ nm$^{-2}$, respectively for $\rho_{core}$ and $\rho_{solv}$, are typical for a hydrocarbon chain in sophorolipids[11] and for water. $\rho_{shell}$ accounts for the carbohydrate moieties, water and counterions, and it is always a variable parameter, although it should be contained between the hydrated and dehydrated sophorose, that is between 10.0 and 14.0 x 10$^{-4}$ nm$^{-2}$.[11] If the overall quality of the fit can be followed by the classical $\chi^2$ evolution test, a realistic estimation of the error on the final values is always difficult, although an error of ±10% is not outrageous. In our fitting strategy, the starting best-fit parameters are determined on the basis of previous work;[10] $R$, $T$ and $\rho_{shell}$ are then varied keeping $\frac{R_1}{R} = \frac{T_1}{T} = 1$. The fit is then refined by varying $\frac{R_1}{R}$ and $\frac{T_1}{T}$ independently.

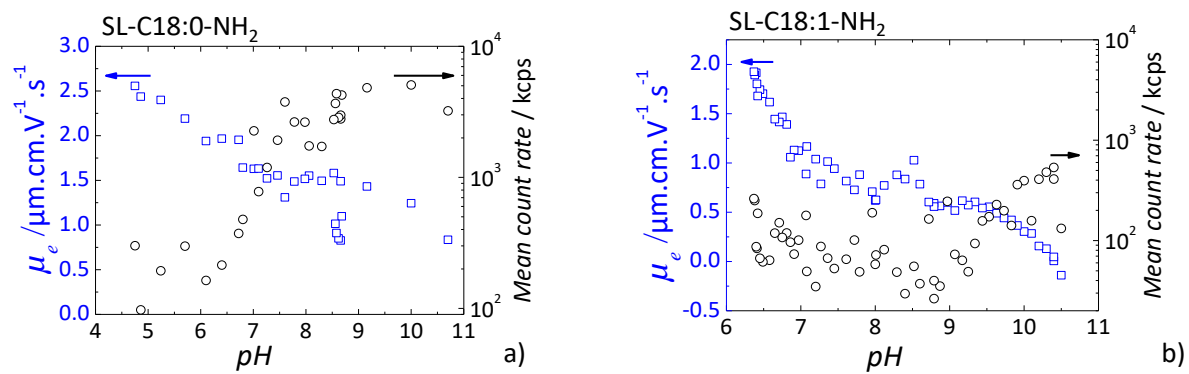

**Figure S 6** – Electrophoretic mobility (blue empty squares) and turbidimetric (black empty circles) experiments performed on a) SL-C18:0-NH$_2$ (C= 2 mg/mL) and b) SL-C18:1-NH$_2$ (C= 5 mg/mL) solutions. The experimental conditions are given in the materials and method section.

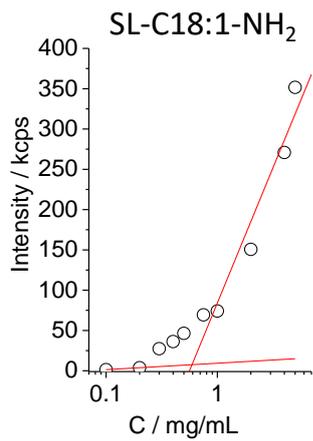

CMC= 0.5 ± 0.2 mg/mL

**Figure S 7 – Critical micelle concentration study of aminyl sophorolipid SL-C18:1-NH$_2$ using static light scattering at constant shutter opening.**

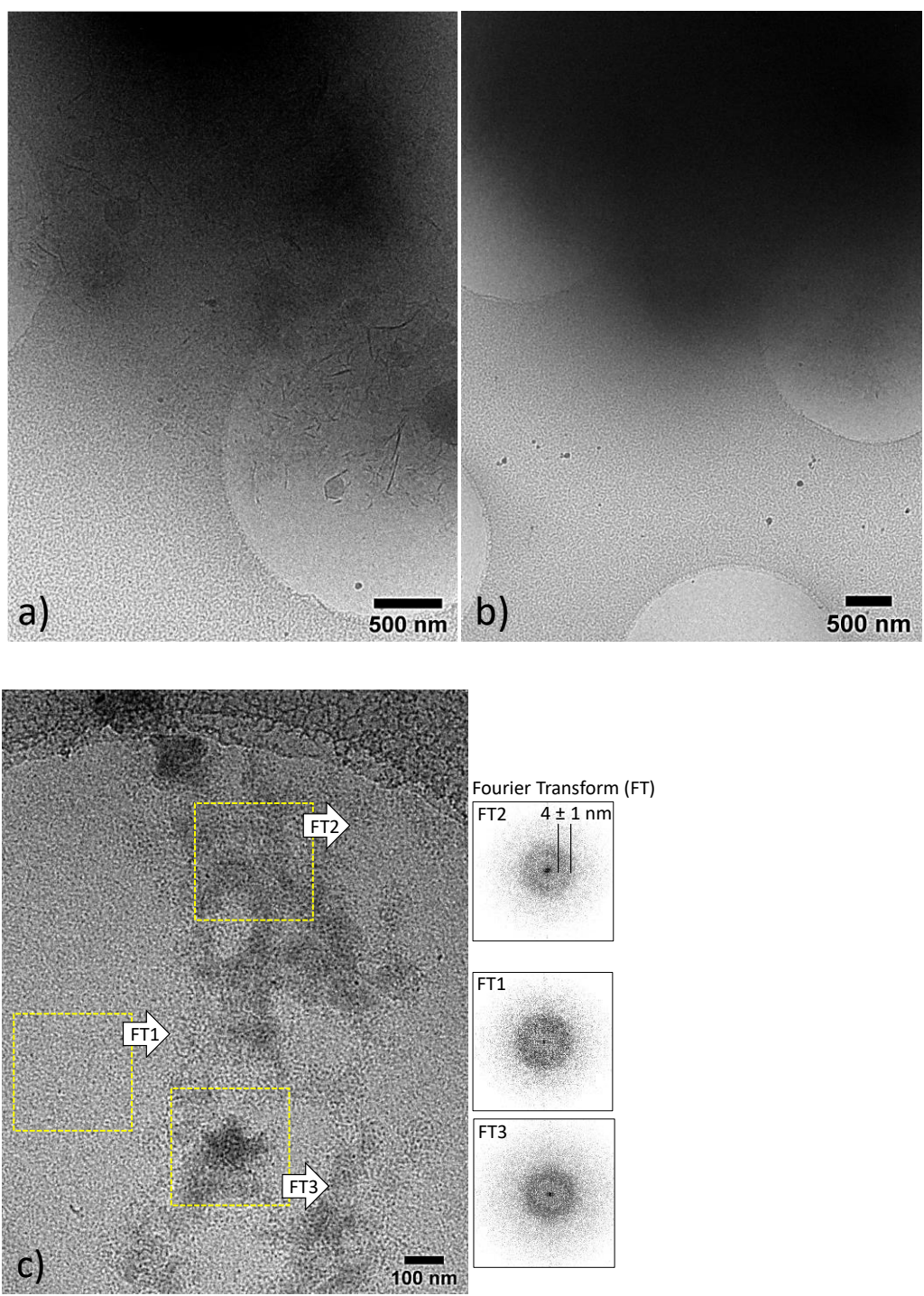

**Figure S 8** – Highlight of the micellar region in cryo-TEM experiments performed on a SL-C18:1-NH$_2$ solution at C= 0.5 wt% and pH 4. a-b) Typical visual look of cloudy regions corresponding to aggregated platelets. These are sitting on top of the holey carbon grid, where holes contain vitrified water. c) Close-up of a vitrified aqueous hole in another region of the same grid. Micellar aggregates are visible within the vitrified layer of water. The Fourier Tranform (FT) corresponding to the yellow highlights are given on the right-hand side. The broad scattering ring identifies a correlation distance of 4 ± 1 nm, in agreement with the micellar diameter measured by SAXS.

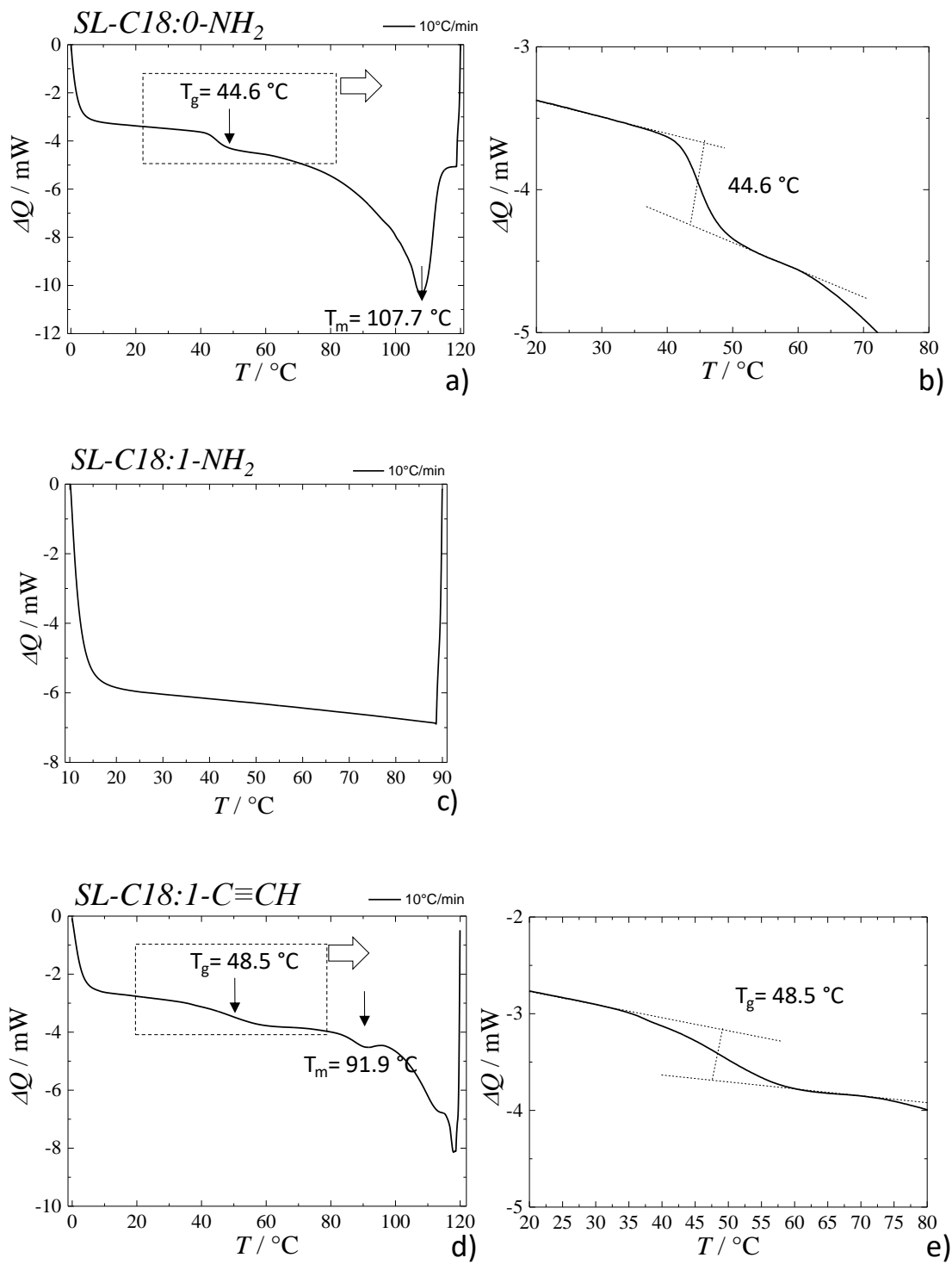

**Figure S 9** – DSC experiments performed on SL-C18:0-NH₂, SL-C18:1-NH₂ and SL-C18:1-C≡CH powder samples at heating rate of 10°C/min. Error on the value of $T_g$ is estimated to be ± 0.5°C and it is related to the uncertanty associated to the tangets method. The value of $T_m$ is determined by fitting the profile with a Lorentzian peak.